\documentclass[conference]{IEEEtran}
\IEEEoverridecommandlockouts

\usepackage[moderate,tracking=normal]{savetrees}
\usepackage[ruled,linesnumbered]{algorithm2e}
\usepackage{makecell}
\usepackage{enumitem}
\usepackage{bm}
\usepackage{multirow}
\usepackage{cite}
\usepackage{amsmath,amssymb,amsfonts}
\usepackage{graphicx}
\usepackage{textcomp}
\usepackage{xcolor}
\usepackage{hyperref}
\usepackage{titlesec}
\usepackage{balance}

\graphicspath{{figs/}}
\def\BibTeX{{\rm B\kern-.05em{\sc i\kern-.025em b}\kern-.08em
    T\kern-.1667em\lower.7ex\hbox{E}\kern-.125emX}}

\newcommand{\tabwidth}{.95\columnwidth}

\titlespacing*{\section}{0pt}{4pt}{4pt}
\titlespacing*{\subsection}{0pt}{4pt}{4pt}
\titlespacing*{\subsubsection}{0pt}{4pt}{4pt}
\setlist{nosep}
\setlist[itemize]{topsep=0pt,partopsep=0pt}
\setlist[enumerate]{topsep=0pt,partopsep=0pt}
\newcommand{\Queue}[1]{\ensuremath{\mathcal{Q}_{\mathrm{#1}}}}

\newcommand{\QueueRed}[1]{\textcolor{red}{\ensuremath{\mathcal{Q}_{\mathrm{#1}}}}}
\newcommand{\QueueBlue}[1]{\textcolor{blue}{\ensuremath{\mathcal{Q}_{\mathrm{#1}}}}}
\SetKwInput{KwLaunch}{Launch}
\SetKwFor{ForPar}{parfor}{do}{end}

\newcommand{\BlockComment}[1]{\textbf{/* #1 */}}

\newcommand{\algfontsize}{\footnotesize}
\def\wideBarScale{.9}

\begin{document}

\title{$g$MAGNUS: Fast SpGEMM on GPUs for Irregular Matrices via Hierarchical Multisplit}

\author{
\IEEEauthorblockN{Jordi Wolfson-Pou}
\IEEEauthorblockA{\textit{Intel} \\
Santa Clara, USA \\
jordi.wolfson-pou@intel.com}
\and
\IEEEauthorblockN{Ahmed Helal}
\IEEEauthorblockA{\textit{Intel} \\
Santa Clara, USA \\
ahmed.helal@intel.com}
\and
\IEEEauthorblockN{Fabrizio Petrini}
\IEEEauthorblockA{\textit{Intel} \\
Santa Clara, USA \\
fabrizio.petrini@intel.com}
}

\maketitle

\begin{abstract}
We present $g$MAGNUS, a novel algorithm for sparse matrix-matrix multiplication (SpGEMM) of irregular matrices on GPUs. Such matrices often contain many \emph{heavy rows}, those with large intermediate products that force local memory accumulators to spill to global memory.  $g$MAGNUS addresses this by computing an intra-row reordering of intermediate products, subdividing heavy rows into independent chunks that can be accumulated completely in local memory. This reordering uses novel outer product and hierarchical multisplit operations. The algorithm is input- and system-aware, automatically determining the number of chunks and multisplit levels based on the input matrix dimensions and local memory size.  Experimental results on two extensive datasets show that $g$MAGNUS achieves a geometric-mean speedup of $1.81$--$7.62\times$ over five leading algorithms (including MKL and cuSPARSE) on Intel Ponte Vecchio and NVIDIA H200. Additionally, the core kernels of $g$MAGNUS are evaluated, achieving near-peak performance compared to their theoretical upper bound.
\end{abstract}

\begin{IEEEkeywords}
SpGEMM, GPUs, Gustavson, Outer product
\end{IEEEkeywords}

\section{Introduction}
\sloppy The sparse general matrix-matrix multiplication 
(SpGEMM), which multiplies two sparse matrices $A$ and $B$ to produce a sparse matrix $C$, is critical to the performance of a wide range of applications~\cite{spgemmSurvey}, including genome assembly~\cite{genome1,genome2,bella}, machine learning~\cite{pca,mcl,dnn,HipMCL,hoefler}, algebraic multigrid~\cite{amg1,amg2}, and graph analytics~\cite{bfs,subgraph,tricount1,tricount2,color,tricount3}.
Although GPUs are widely used to accelerate SpGEMM,
their performance is often limited by the difficulty of mapping sparse structures onto the massive compute resources of GPUs. 
The challenge becomes more pronounced as matrices grow larger (both in the dimensions and nonzeros per row) and more structurally irregular, introducing high levels of imbalance and uncoalesced accesses to global memory.

Gustavson's method is the basis of most SpGEMM algorithms, processing matrices $A$ and $B$ in row-major order. 
Nonzero elements in each row of $A$ are multiplied with corresponding rows of $B$ to form intermediate products, which are then summed in an accumulator to produce the final row of $C$. 
The most popular GPU accumulator is a hash map with column indices as keys. 
Upon insertion, new and existing values with duplicate keys are summed.
Once all nonzeros are processed, the hash map is sorted to produce the output row.
The standard approach keeps accumulators in local memory, where irregular accesses are much faster than global memory. This suffices for highly structured or small sparse matrices with sparse intermediate products. 
Banded matrices exemplify this case due to their nearest-neighbor structure, which produces small, localized intermediate products that can often be accumulated by a local-memory dense accumulator.
Random power-law matrices are far more challenging: input row sizes vary widely, and some rows produce large intermediate products distributed across the column space.

Scaling up these highly irregular matrices amplifies this issue, creating many \emph{heavy rows} where intermediate products exceed the local-memory hash map capacity, forcing global memory fallbacks.
One approach uses global spillover queues when local memory capacity is reached, 
requiring an unbounded number of passes of the queue, creating load imbalance when some work groups perform many more passes than others~\cite{anh}. 
More recent algorithms use backup global memory accumulators accessed only when local memory overflows, creating uncoalesced accesses to global memory~\cite{OpSparse,spECK}.
Another approach is to sort and merge the intermediate products, which can be far too expensive for large numbers of intermediate elements.


We present $g$MAGNUS (\textbf{M}atrix \textbf{A}lgebra for \textbf{G}igantic \textbf{NU}merical \textbf{S}ystems on GPUs), inspired by prior CPU work~\cite{magnus}, designed for massive irregular matrices with heavy rows.
In the CPU approach, the goal was to maximize L2 cache reuse during accumulation.
Each thread processed a distinct set of heavy rows, performing an outer product and an intra-row reordering of their intermediate products independently of other threads.
This reordering introduced sufficient locality for the accumulator to fit in the L2 cache.
The goal of $g$MAGNUS is to generate enough locality to perform accumulation entirely in local memory.

Achieving this on GPUs requires a markedly different design, as work groups and threads must coordinate closely to maintain load balance and memory coalescing.
To this end, $g$MAGNUS first uses an outer product to construct the global intermediate matrix $\hat{C}$, then applies a novel hierarchical multisplit~\cite{gpu-multisplit} to reorder each row of $\hat{C}$ into chunks that form the rows of $\hat{C}_{\text{reord}}$.
This hierarchical design supports an unbounded number of chunks, so it scales to matrices of any size by further subdividing rows through a small number of passes over $\hat{C}_{\text{reord}}$.
As a result, the column range of each row in $\hat{C}_{\text{reord}}$ is much smaller than that of $C$, allowing a dense accumulator to fit entirely in local memory and avoiding fallback to global-memory accumulation.
Finally, $g$MAGNUS is input- and system-aware, automatically selecting the number of chunks and hierarchy levels from the input matrix dimensions and available local memory.

We implemented $g$MAGNUS in SYCL and CUDA, and present experimental results on both Intel Ponte Vecchio and NVIDIA H200 GPUs.
Five SpGEMM algorithms are used as baselines:
Intel MKL~\cite{mkl}, Kokkos~\cite{kokkos,kokkos2}, NVIDIA cuSPARSE~\cite{cuSPARSE}, TileSpGEMM~\cite{tileSpGEMM}, and OpSparse~\cite{OpSparse}.
We consider two datasets: the SuiteSparse matrix collection~\cite{suitesparse} and recursive power-law matrices (RMats) generated by PaRMAT~\cite{PaRMAT}.
Across all GPUs and datasets, $g$MAGNUS is faster than all baselines in terms of the geometric-mean speedup, with higher speedups for massive matrices that have large dimensions and intermediate products, such as large-scale RMat matrices.
Additionally, we compare the core kernels of $g$MAGNUS (outer product and multisplit) to their ``speed of light'' performance, i.e., their theoretical peak performance.
The results show that our core kernels achieve near-peak performance.
The outer product often achieves 60-80\% of the peak for large test matrices.
The multisplit kernel, which has been studied for more regular applications, attains $40$--$50\%$ of the peak, consistent with more extensive studies~\cite{gpu-multisplit}.

\section{Background}\label{sec:background}
\subsection{Notation and Definitions}
Let $A$ be a sparse matrix, with $A.n$, $A.m$, and $A.\mathit{nnz}$ denoting its numbers of rows, columns, and nonzeros, respectively.
We use direct indexing: $A[i,:]$, $A[:,j]$, and $A[i,j]$ to denote row $i$, column $j$, and element $(i,j)$, respectively.
Sparse matrices are stored in the widely used compressed sparse row (CSR) format, where $A.\mathit{col}$, $A.\mathit{val}$, and $A.\mathit{rowPtr}$ store the column indices, values, and row pointers, and have sizes $A.\mathit{nnz}$, $A.\mathit{nnz}$, and $A.n+1$, respectively.
When indexing into a CSR matrix, row pointers determine the accessed range, e.g., $A[i,:]$ denotes elements $[A.\mathit{rowPtr}[i]:A.\mathit{rowPtr}[i+1])$ of $A.\mathit{col}$ and $A.\mathit{val}$.
Vectors are denoted in lowercase (e.g., $x$, $v$), with $x.n$ denoting the size.
Subarrays and submatrices are indexed as $[\mathit{start}:\mathit{end})$, so $x[\mathit{start}:\mathit{end})$ denotes the consecutive elements $\{x[\mathit{start}], \dots, x[\mathit{end}-1]\}$.
A masked subarray $x[v]$ denotes the elements $\{x[v[0]], \dots, x[v[v.n-1]]\}$.
Similarly, $A[v,:]$ and $A[:,v]$ denote the submatrices formed by selecting the rows or columns indexed by $v$.
On GPUs, we follow SYCL terminology: \emph{work group} (\emph{thread block}), \emph{subgroup}, and \emph{local memory} (shared memory).
A \emph{device queue} denotes a SYCL queue (CUDA stream).
We use \emph{parfor} to denote a GPU-style strided parallel loop over $i\in [\mathit{start},\mathit{end})$.

\subsection{Gustavson's Method}\label{sec:gustav}
Gustavson's 
method~\cite{gustavson} is the basis for most GPU SpGEMM algorithms, and can be written as:
\begin{equation}
    C[i,\; :] = \sum_{j = 0}^{A.m-1} A[i,j]\times B[j,\; :].
    \label{equ:gustav}
\end{equation}
That is, the nonzeros in row $i$ of $A$ scale the corresponding rows of $B$, and the resulting scaled rows are summed to produce row $i$ of $C$.
Because the rows of $C$ are independent, Gustavson's method naturally parallelizes over rows.
\autoref{alg:gustav} shows a CSR version of the \emph{numeric phase} of Gustavson's method.
The numeric phase populates $C.\mathit{col}$ and $C.\mathit{val}$, while the \emph{symbolic phase}, performed beforehand, 
computes $C.\mathit{rowPtr}$ and $C.\mathit{nnz}$ (see~\cite{spgemmSurvey} for more on \emph{size prediction}).
To combine the \emph{intermediate product} of a row, that is, the set of scaled rows $A[i,j]\times B[j,\; :]$, an accumulator $\mathit{accumBuff}$ sums the contributions on the fly.
After accumulation, a pass over $\mathit{accumBuff}$ sorts and writes the row to $C$; in some implementations, the entire matrix $C$ is sorted afterwards.
In typical parallel GPU implementations, the outer loop is partitioned across work groups and the inner loops across threads within a work group.

\begin{algorithm}[tbp]
    \algfontsize
    \caption{Gustavson's Method}\label{alg:gustav}
    \KwIn{$\bm{A}$, $\bm{B}$, $\bm{C.\mathit{rowPtr}}$}
    \KwOut{$\bm{C.\mathit{col}}$, $\bm{C.\mathit{val}}$}
    \SetCommentSty{emph}
    \DontPrintSemicolon
    \For{$i \in [0:A.n)$}{
        \For{$j \in [A.\mathit{rowPtr}[i]:A.\mathit{rowPtr}[i+1])$}{
            \For{$k \in [B.\mathit{rowPtr}[A.\mathit{col}[j]]:B.\mathit{rowPtr}[A.\mathit{col}[j]+1])$}{
                $\mathit{accumBuff}[B.\mathit{col}[k]]$ \texttt{+=} $A.\mathit{val}[j] \times  B.\mathit{val}[k]$\;
            }
        }
        $C[i,:] \gets \texttt{SortAndWrite}(\mathit{accumBuff})$\;
    }
\end{algorithm}

A key consideration in this on-the-fly accumulation is whether the accumulator $\mathit{accumBuff}$ fits in local memory, which is critical to efficiently handle the irregular accesses that arise during accumulation.
The most common accumulator designs are dense arrays, hash maps, and sort-based algorithms.
For ``well-behaved'' matrices, such as banded matrices, structural locality yields small, localized intermediate products, allowing a shortened dense accumulator whose size is proportional to the bandwidth rather than $C.m$.
For more general matrices, hash maps and sort-based accumulators are more robust, since a dense accumulator requires storage proportional to $C.m$, which is often prohibitively expensive.
However, for highly irregular matrices, such as random power-law matrices, many rows produce intermediate products that exceed local memory capacity, forcing all three approaches to use global memory and requiring more complex algorithms.
We refer to such rows as \emph{heavy rows}, and to the remaining rows as \emph{light rows}.

\subsection{ESC and The Outer Product}
In some variants of Gustavson’s method, such as expand--sort--contract (ESC)~\cite{ESC}, intermediate products are explicitly stored.
A first pass expands these products to form the intermediate matrix $\hat{C}$, and a subsequent accumulation pass merges rows of $\hat{C}$ to produce $C$.
Alternative approaches adopt an outer-product formulation~\cite{pbSpGEMM,outerspace,magnus,blockReorg,sparch,outerProdSocialNetworks}, defined as:
\begin{equation}
C = \sum_{i=0}^{A.m-1} A[:,i] \otimes B[i,:],
\label{equ:outer}
\end{equation}
where each rank-1 update (or \emph{slice}) is given by the outer product of column $i$ of $A$ and row $i$ of $B$.
This formulation mitigates a key drawback of Gustavson’s method, namely uncoalesced accesses and limited reuse in $B$, at the cost of explicitly constructing $\hat{C}$.
\autoref{alg:outer} shows a simple parallel version of the outer product.
The first pass performs the construction of $\hat{C}$ using a parallel loop over the slices.
The two inner loops illustrate the reuse of $B$, where row $i$ of $B$ is reused for each nonzero in column $i$ of $A$.
Since each slice modifies rows of $\hat{C}$ corresponding to the row indices in column $i$ of $A$, an atomic fetch-and-add allows for multiple slices to write to the same row.

\begin{algorithm}[tbp]
    \algfontsize
    \caption{Outer Product}\label{alg:outer}
    \KwIn{$A$, $B$, $C.\mathit{rowPtr}$}
    \KwOut{$C.\mathit{col}$, $C.\mathit{val}$}
    \SetCommentSty{emph}
    \DontPrintSemicolon
    \ForPar{$i \in [0:A.m)$}{
        \For{$j \in [A_{\text{CSC}}.colPtr[i]:A_{\text{CSC}}.colPtr[i+1])$}{
            \For{$k \in [B.\mathit{rowPtr}[i],\; B.rowPtr[i+1])$}{
                $\mathit{dest} \gets \texttt{AtomicAdd}(\hat{C}.\mathit{rowPtr}[A.\mathit{row}[j]+1],\; 1)$\;
                $\hat{C}.\mathit{col}[\mathit{dest}]\gets B.\mathit{col}[k]$\;
                $\hat{C}.val[\mathit{dest}]\gets A.val[j]\times B.val[k]$\;
            }
        }
    }
    \ForPar{$i \in [0,\; \bm{C.n})$}{
        $\bm{C}[i,:] \gets \texttt{Accum}(\bm{\hat{C}}[i,:])$\;
    }
\end{algorithm}

While outer product improves reuse and mitigates the uncoalesced reads in Gustavson’s method, it introduces new challenges on GPUs.
First, load balancing remains critical and requires careful distribution of work across all loop levels.
Second, atomics can be avoided by expanding $\hat{C}$ in coordinate format, but the resulting triplets must be sorted during the final accumulation pass, which is prohibitive for large intermediate products.
Third, achieving coalesced writes to $\hat{C}$ further requires careful partitioning of loops so that consecutive threads write to consecutive addresses.
Finally, outer product 
primarily addresses inefficiencies in reading the input matrices, but the challenge of accumulating heavy rows still remains.

\subsection{Related Work}
We focus on recent GPU SpGEMM algorithms most relevant to $g$MAGNUS. A comprehensive survey is available in\cite{spgemmSurvey}. 

Most prior work has focused on optimizing local-memory accesses and improving accumulator load balancing within Gustavson's method~\cite{nsparse,anh,spECK,OpSparse,HSMU-SpGEMM}.
A common strategy is to use one or more accumulator types and \emph{bin} rows by size.
Smaller work groups process smaller rows, while rows that exceed the local-memory capacity of the largest work group either require multiple passes~\cite{anh,spECK} or fall back to a global-memory accumulator~\cite{nsparse,OpSparse,kokkos}.
Hybrid accumulators are also common, for example using hash maps for sparser rows or tiles and dense accumulators otherwise~\cite{tileSpGEMM,spECK}, or using different accumulator types in the symbolic and numeric phases~\cite{HSMU-SpGEMM}.

The 
ESC-based 
methods~\cite{ESC,ESC2,cusp,AC-SpGEMM,bhSparse,iterMerge} are generally less competitive than Gustavson-based approaches because they explicitly generate and sort the intermediate product in global memory, which is expensive.
Recent ESC variants reduce the cost of global sorting by sorting smaller subarrays in local memory~\cite{AC-SpGEMM}, but this requires additional passes over global memory to generate and merge those subarrays.

Other work explores storage formats beyond CSR~\cite{tSparse,tileSpGEMM}, of which
TileSpGEMM~\cite{tileSpGEMM} is a leading approach.
Inspired by tiled dense GEMM, TileSpGEMM converts the input matrices into $16\times16$ tiles, storing only nonzero tiles, which improves locality in both the input matrices and the accumulator.
However, this introduces a substantial CSR-to-tile setup cost, and the output matrix $C$ must later be converted from tiled form back to CSR.
\section{MAGNUS}\label{sec:magnus}
\subsection{Overview}
$g$MAGNUS addresses heavy rows in large irregular matrices by performing an intra-row reordering of the intermediate matrix $\hat{C}$ (henceforth, $\hat{C}$ denotes the intermediate matrix generated only by the heavy rows).
It splits rows into independent chunks to produce a second intermediate matrix, $\hat{C}_{\text{reord}}$, whose structure both enables local-memory accumulation and exposes additional intra-row parallelism.
This reordering is implemented through an outer product and a novel hierarchical multisplit operation, both optimized for load balancing and coalesced memory accesses.
The final accumulation step uses a local-memory hybrid strategy: hash map accumulation for light rows of $C$ and $\hat{C}_{\text{reord}}$, and dense accumulation for heavy rows of $\hat{C}_{\text{reord}}$.
Our approach is input- and system-aware, choosing the number of chunks per row based on $C.m$ and the local-memory capacity.
These parameters determine the minimum number of chunks needed for dense accumulators to fit in local memory (\autoref{sec:magnus_params} shows how the number of chunks is determined at runtime).
Because it explicitly stores intermediate products, $g$MAGNUS is similar in spirit to ESC-based algorithms~\cite{ESC}.
However, by expanding only the intermediate products of heavy rows, it requires substantially less memory than prior ESC approaches.

\begin{figure}[tbp]
\centering
\begin{tabular}{c}
\includegraphics[width=.85\linewidth]{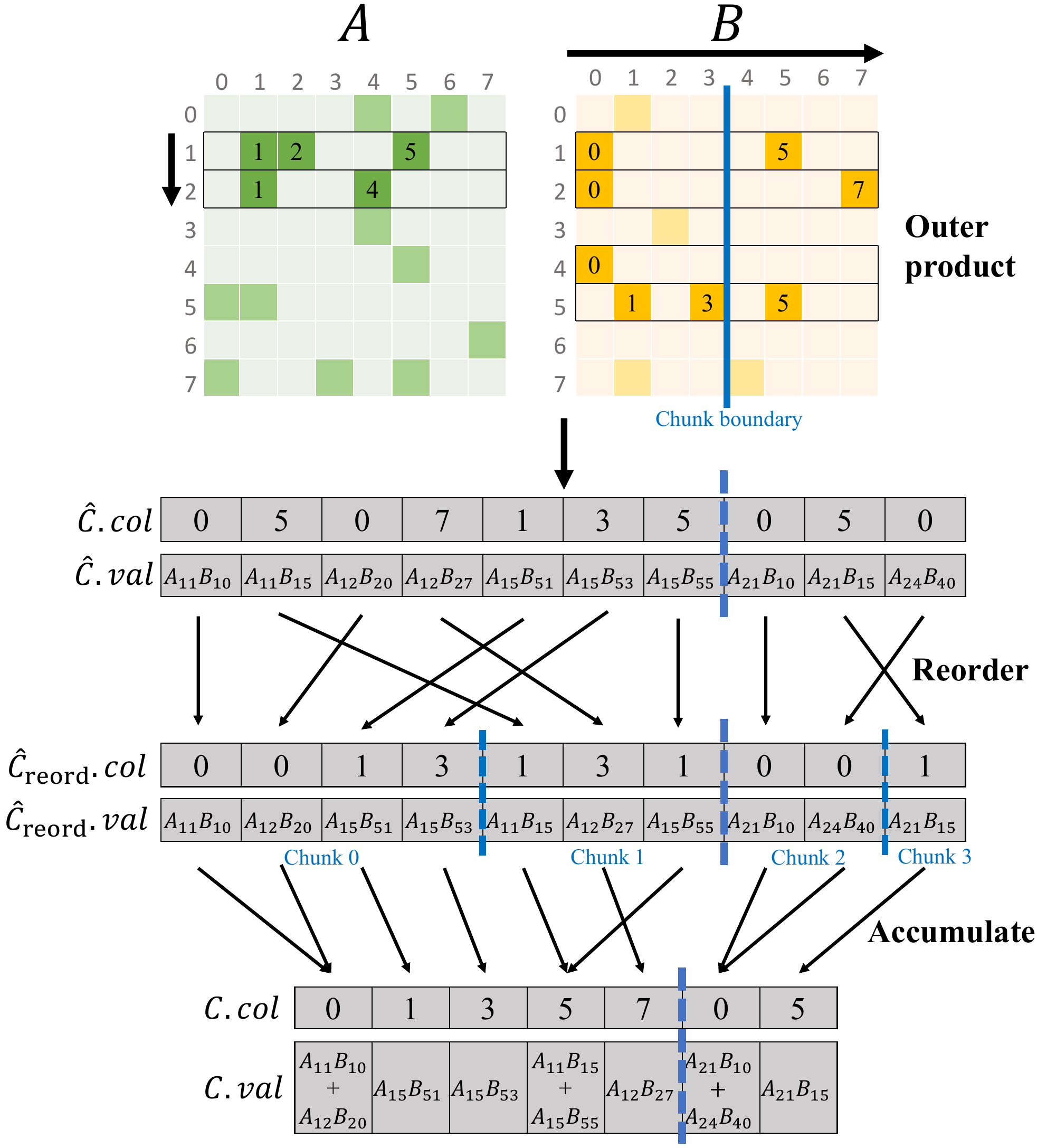} 
\end{tabular}
\caption{Example of processing of heavy rows by $g$MAGNUS, where two chunks and one level are used to compute rows 1 and 2 of $C$.  The dashed lines represent the row pointers, whereas the column and value arrays are explicitly written.}
\label{fig:magnus_example}
\end{figure}

Our optimized multisplit kernel performs a range-based multisplit operation, permuting a set of key-value pairs into \emph{chunks} (also referred to as \emph{buckets} or \emph{bins}), where elements belonging to the same chunk are stored contiguously.
The set of key-value pairs are rows of $\hat{C}$ and the chunks are rows of $\hat{C}_{\text{reord}}$, e.g., if $\hat{C}$ stores two heavy rows that must be split into two chunks each, $\hat{C}_{\text{reord}}$ will have 4 rows, as shown in \autoref{fig:magnus_example}.
While $\hat{C}_{\text{reord}}$ has twice as many rows as $\hat{C}$, it has half as many columns.
The reduced column range enables local memory-only dense accumulation, while the increased number of rows exposes additional parallelism by effectively splitting each row of $\hat{C}$ across multiple work groups.
Within a chunk, elements are unordered and processed in later stages by the accumulator.
To construct the rows of $\hat{C}_{\text{reord}}$, column indices are shifted into their local chunk range.
in \autoref{fig:magnus_example}, chunks 1 and 3 shift indices from $[4,8)$ to $[0,4)$.
During the final write to $C$, these indices are shifted back to their original range.


\autoref{alg:magnus} shows the end-to-end $g$MAGNUS algorithm, where all inputs and outputs reside in device memory.
The notation $\langle y_1,y_2,\dots \rangle \gets \texttt{Function}\langle\Queue{}\rangle(x_1,x_2,\dots)$ denotes a function that launches one or more kernels on device queue $\Queue{}$.
Light and heavy rows are processed concurrently on separate asynchronous device queues.
In the setup phase, a preprocessing step queries the device and computes the heavy-row threshold $\tau$, the number of chunks per row per level $v_{\text{chunks}}$, and the number of levels $v_{\text{chunks}}.n$
(as detailed later in \autoref{sec:magnus_params}).
$\tau$ is derived from the local memory size in bytes, $s_{\text{LM}}$, which determines the maximum accumulator capacity.
This $O(1)$ computation overlaps with \texttt{InterOffsets()}, which computes per-row intermediate product sizes and generates $\hat{C}.\mathit{rowPtr}$ using a device-wide prefix sum. 
The intermediate product sizes together with $\tau$ are used to partition rows into light and heavy groups using 
vendor primitives, such as \texttt{cub::DevicePartition::Flagged()} in CUDA.
The remaining steps execute on the two concurrent queues for light and heavy rows.

\begin{algorithm}[tbp]
    \small
    \SetCommentSty{emph}
    \DontPrintSemicolon
    \caption{$g$MAGNUS}\label{alg:magnus}
    \KwIn{$A$, $B$}
    \KwOut{$C$}
    \BlockComment{Setup phase}\;
    $\langle \tau,\; v_{\text{chunks}} \rangle \gets \texttt{GetMagnusParams}(C.m,\; s_{\text{LM}})$\;
    $\hat{C}.\mathit{rowPtr} \gets \texttt{InterOffsets}\langle\Queue{main}\rangle(A,\; B)$\;
    $\langle x_{\text{heavy}},\; x_{\text{light}}\rangle \gets \texttt{ThreshPartition}\langle\Queue{main}\rangle(\hat{C}.rowPtr,\; \tau)$\;
    Synchronize()\;

    \BlockComment{Outer product}\;
    $\langle \tilde{A}_{\text{CSC}}, \tilde{B} \rangle \gets \texttt{MaskedCsr2Csc}\langle\QueueBlue{heavy}\rangle(A[x_{\text{heavy}},:], B)$\;\label{alg:magnus:outer_init}
    $\hat{C} \gets \texttt{OuterProduct}\langle\QueueBlue{heavy}\rangle(\tilde{A}_{\text{CSC}},\; \tilde{B})$\;\label{alg:magnus:outer_final}
    
    \BlockComment{Reorder}\;
    \For{$i \in [0:v_{\text{chunks}}.n)$}{\label{alg:magnus:reord_init}
        $\hat{C}_{\text{reord}}.rowPtr \gets \texttt{Histogram}\langle\QueueBlue{heavy}\rangle(\hat{C},\; v_{\text{chunks}}[i])$\;

        \If{$i < v_{\text{chunks}}.n-1$}{\label{alg:magnus:last_level}
            $\langle \hat{x}_{\text{heavy}},\; \hat{x}_{\text{light}}\rangle \gets \texttt{ThreshPartition}\langle\QueueBlue{heavy}\rangle(\hat{C}_{\text{reord}}.rowPtr,\; \tau)$\;
        }
        
        $\hat{C}_{\text{reord}} \gets \texttt{Multisplit}\langle\QueueBlue{heavy}\rangle(\hat{C},\hat{C}_{\text{reord}},\;\; v_{\text{chunks}}[i])$\;
        
        \If{$\hat{x}_{\text{heavy}}.n$ == $0$}{\label{alg:magnus:no_more_heavy}
            break\;
        }
        $\hat{C} \gets \hat{C}_{\text{reord}}[\hat{x}_{\text{heavy}},:]$\;
    }\label{alg:magnus:reord_final}
    $C.\mathit{rowPtr}[x_{\text{heavy}}] \gets \texttt{Symbolic}\langle\QueueBlue{heavy}\rangle(\hat{C}_{\text{reord}})$\;
    $C.\mathit{rowPtr}[x_{\text{light}}] \gets \texttt{Symbolic}\langle\QueueRed{light}\rangle(A[x_{\text{light}},:], B)$\;
    Synchronize()\;
    $C.\mathit{rowPtr} \gets \texttt{PrefixSum}\langle\Queue{main}\rangle(C.\mathit{rowPtr})$\;
    Synchronize()\;
    
    $C[x_{\text{heavy}},:] \gets \texttt{Numeric}\langle\QueueBlue{heavy}\rangle(\hat{C}_{\text{reord}})$\;
    $C[x_{\text{light}},:] \gets \texttt{Numeric}\langle\QueueRed{light}\rangle(A[x_{\text{light}},:], B)$\;
    Synchronize()\;
\end{algorithm}


To process heavy rows, we first perform the outer product (lines~\ref{alg:magnus:outer_init}-\ref{alg:magnus:outer_final}), which requires matrix $A$ in CSC format. 
Hence, we carry out a masked CSR-to-CSC conversion using a key-value sorting approach~\cite{amg3}.
Additionally, we form the masked matrix $\tilde{B}$,  which contains the rows of $B$ corresponding to the nonzero columns of $\tilde{A}_{\text{CSC}}$.
The outer product then populates the intermediate matrix $\hat{C}$.

Our hierarchical multisplit 
(shown at lines~\ref{alg:magnus:reord_init}--\ref{alg:magnus:reord_final}) consists of multiple operations.
\texttt{Histogram()} 
counts the number of elements per chunk and performs a device-wide prefix sum to compute $\hat{C}_{\text{reord}}.\mathit{rowPtr}$.
\texttt{Multisplit()} then permutes the entries of $\hat{C}$ to populate $\hat{C}_{\text{reord}}$.
Then we again use $\tau$ to split the remaining chunks into light and heavy groups, since only heavy chunks need further reordering, reducing data volume for subsequent multisplit operations.
If there are only light chunks, we break from the loop (line~\ref{alg:magnus:no_more_heavy}).
At the last level (i.e., when $i$ == $v_{\text{chunks}}.n-1$), $\hat{C}_{\text{reord}}.m$ is small enough for the dense accumulator to fit in local memory, meeting the $g$MAGNUS design goal.
The remaining steps of $g$MAGNUS follow the standard SpGEMM workflow, with accumulation in the symbolic and numeric stages (see \autoref{sec:accum}).

For algorithmic readability, we show explicit gather operations like $\hat{C} \gets \hat{C}_{\text{reord}}[\hat{x}_{\text{heavy}},:]$. 
However, in the implementation, this operation is fused with our GPU kernels.
For example, this gather is actually performed within \texttt{Multisplit()} during the final write to global memory, ensuring heavy chunks are ordered first so that subsequent multisplits operate on contiguous elements for optimal load balancing.
Additionally, the loop over levels uses double buffering: one buffer holds the input $\hat{C}$, produced either by the outer product or by the reordering step from the previous iteration, while the other holds the output $\hat{C}_{\text{reord}}$. After the loop completes, the buffer containing the final input is deallocated.

\subsection{Outer Product Kernel}

\begin{figure}[tbp]
\centering
\begin{tabular}{c}
\includegraphics[width=.85\linewidth]{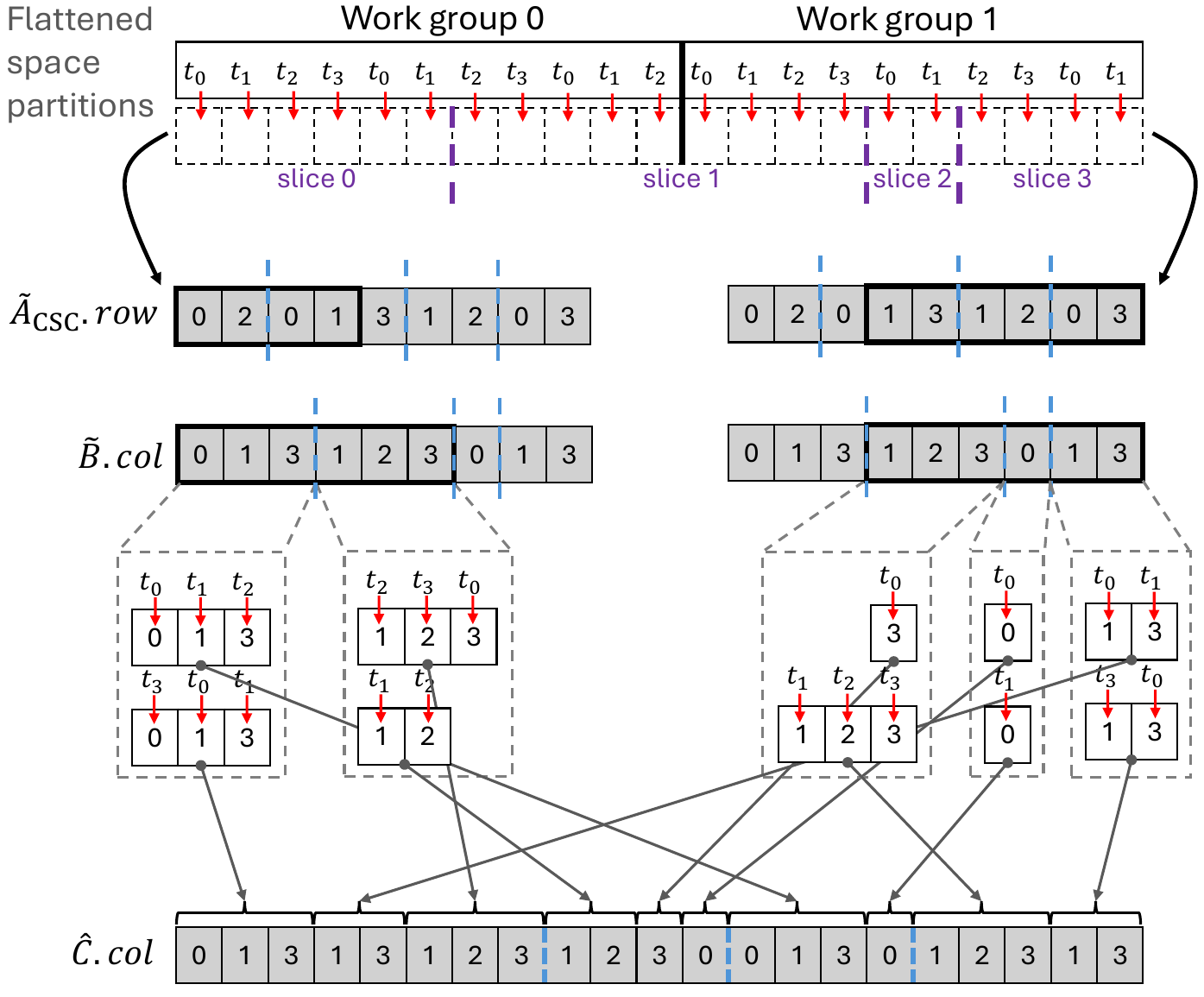} 
\end{tabular}
\vspace*{-3pt}
\caption{Example partitioning of $\tilde{A}_{\text{CSC}}$ and $\tilde{B}$ for the outer product kernel.
The top sequence of elements shows the partitioning of the flattened index space, which maps to the partitioning of $\tilde{A}_{\text{CSC}}$ and $\tilde{B}$ shown below it.
The blue dashed boundary lines represent row pointers.
The elements read by each thread are also shown, as calculated by \texttt{GetEntry()} and \texttt{GetExcess()} in \autoref{alg:magnus_outer}.}
\label{fig:outer_partition_example}
\end{figure}

The outer product kernel ($\texttt{OuterProduct()}$ in \autoref{alg:magnus}) balances memory traffic across work groups and maximizes coalesced accesses.
We use a static partitioning computed locally without any inter-thread coordination.
Each work group reads the same number of $\tilde{A}_{\text{CSC}}$–$\tilde{B}$ element pairs, multiplies them, and writes the results to $\hat{C}$.
We achieve this with a static partitioning of a flattened index space, as shown in the top array of
 \autoref{fig:outer_partition_example}, where two work groups are assigned approximately the same number of slice (input) and intermediate (output) elements.
This flattening unrolls the slices and requires mapping indices in the flattened space back to elements in $\tilde{A}_{\text{CSC}}$ and $\tilde{B}$.
Both work groups 
read consecutive elements of $\tilde{A}_{\text{CSC}}$ and $\tilde{B}$ and write consecutive elements of $\tilde{B}$ to $\hat{C}$.

The mapping from the flattened space to $\tilde{A}_{\text{CSC}}$ and $\tilde{B}$ must handle two important edge cases.
First, elements of $\tilde{A}_{\text{CSC}}$ and $\tilde{B}$ read by different work groups may overlap, for example, see slice 1 (column 1 and row 1 of $\tilde{A}_{\text{CSC}}$ and $\tilde{B}$, respectively).
Second, as a result of overlap, work groups may read partial rows of $\tilde{B}$, as shown in the example where work group 0 reads two of the three elements on its last pass over row 1 of $\tilde{B}$, and work group 1 reads the remaining last element on its first pass over the same row of $\tilde{B}$.

\autoref{alg:magnus_outer} shows our outer product kernel, which has three main steps: (1) local load balancing, (2) updating $\hat{C}.\mathit{rowPtr}$, and (3) populating $\hat{C}$.
In the algorithm, bold symbols are global and symbols with an $\ell$ subscript are local memory arrays.
The mapping $\bm{x_{\text{map}}}$ is derived from the light-heavy row partitioning and maps the heavy rows, which are nonconsecutive in the original ordering, to consecutive rows in $\hat{C}$.
The first launch parameter is the number of work groups, and the second is the work group size.
The slice offsets are computed 
within \texttt{OuterProduct()} as the prefix sum of $\mathit{sliceOffsets}[i+1] = (\tilde{A}_{\text{CSC}}.\mathit{colPtr}[i+1]-\tilde{A}_{\text{CSC}}.\mathit{colPtr}[i])\times (B.\mathit{rowPtr}[i+1]-B.\mathit{rowPtr}[i])$ (the number of elements per slice) for $i\in \{0,1,\dots,A.m-1\}$.

\begin{algorithm}[htbp]
    \setlength{\algomargin}{0pt}
    \algfontsize
    \SetCommentSty{emph}
    \DontPrintSemicolon
    \caption{\texttt{OuterProductKernel}}\label{alg:magnus_outer}
    \KwIn{$\bm{\tilde{A}_{\text{CSC}}},\; \bm{\tilde{B}},\; \bm{\mathit{sliceOffsets}},\; \bm{x_{\text{map}}},\;n_{\text{groupElems}}$}
    \KwOut{$\bm{\hat{C}}$}
    \KwLaunch{$\lceil \frac{\hat{C}.nnz}{n_{\text{groupElems}}} \rceil$, $n_{\text{workGroup}}$}
    \BlockComment{Local load balancing}\;
    $\mathit{groupRange} \gets \texttt{GroupRange}\label{alg:magnus_outer:step1_init}(n_{\text{groupElems}})$\;
    $\mathit{sliceRange} \gets \texttt{Search}(\bm{\mathit{sliceOffsets}}, \; \mathit{groupRange})$\;
    $\mathit{aRange} \gets \texttt{ARange}(\mathit{groupRange},\dots)$\;\label{alg:magnus_outer:step1_final}
    \BlockComment{Update $\bm{\hat{C}.rowPtr}$}\;
    $\mathit{slice} \gets \mathit{sliceRange.start}$\;
    \ForPar{$i \in [\mathit{aRange.start}:\mathit{aRange.end})$}{\label{alg:magnus_outer:step2_init}
        \While{$i \ge \bm{\tilde{A}_{\text{CSC}}.colPtr}[slice+1]$}{
            $slice \gets slice + 1$\;
        }
        $\mathit{rowNnzB} \gets \bm{\tilde{B}.rowPtr}[slice+1]-\bm{\tilde{B}.rowPtr}[slice]$\;
        $excess \gets \texttt{GetExcess}(i,\;slice,\;\dots)$\;
        $\mathit{rowNnzB} \gets \mathit{rowNnzB} - (\mathit{excess.first} + \mathit{excess.last})$\;
        $\mathit{offsetsLocal}_\ell[i-\mathit{aRange.start}]\gets \texttt{AtomicAdd}(\bm{\hat{C}.rowPtr}[\bm{x_{\text{map}}[\tilde{A}_{\text{CSC}}.row}[i]]+1],\; \mathit{rowNnzB})$\;
    }\label{alg:magnus_outer:step2_final}
    \texttt{SyncThreads()}\;
    \BlockComment{Populate $\bm{\hat{C}.col}$ and $\bm{\hat{C}.val}$}\;
    $\mathit{slice} \gets \mathit{sliceRange.start}$\;
    \ForPar{$i \in [\mathit{groupRange.start}:\mathit{groupRange.end})$}{\label{alg:magnus_outer:step3_init}
        \While{$i \ge \bm{\mathit{sliceOffsets}}[\mathit{slice}+1]$}{
            $slice \gets slice + 1$\;
        }
        \BlockComment{Compute read locations of $\tilde{A}_{\text{CSC}}$ and $\tilde{B}$}\;
        $\langle \mathit{entryA}, \mathit{entryB}\rangle \gets \mathit{GetABEntries}(i,\; slice,\; \dots)$\;
        \BlockComment{Retrieve column and value data from $\bm{\tilde{A}_{\text{CSC}}}$ and $\bm{\tilde{B}}$}\;
        $colB \gets \bm{\tilde{B}.col}[\bm{\tilde{B}.rowPtr}[slice]+entryB]$\;
        $valB \gets \bm{\tilde{B}.val}[\bm{\tilde{B}.rowPtr}[slice]+entryB]$\;
        $valA \gets \bm{\tilde{A}_{\text{CSC}}.val}[\bm{\tilde{A}_{\text{CSC}}.colPtr}[slice]+entryA]$\;
        \BlockComment{Compute write location in $\bm{\hat{C}}$}\;
        $excess \gets \texttt{GetExcess}(i, slice, \dots)$\;
        $dest \gets \mathit{offsetsLocal}_\ell[\bm{\tilde{A}_{\text{CSC}}.colPtr}[slice]-aRange.start+entryA]+entryB-excess.first$\;
        \BlockComment{Write to $\bm{\hat{C}}$}\;
        $\bm{\hat{C}.col}[dest] \gets colB$\;
        $\bm{\hat{C}.val}[dest] \gets valA\times valB$\;
    }\label{alg:magnus_outer:step3_final}
\end{algorithm}

Local load balancing (lines~\ref{alg:magnus_outer:step1_init}--\ref{alg:magnus_outer:step1_final}) maps work group partitions to their starting elements in the slice offsets, $\tilde{A}_{\text{CSC}}$, and $\tilde{B}$, where work group $i$ processes elements $[i*n_{\text{groupElems}}, (i+1)*n_{\text{groupElems}})$ in the flattened space  (\autoref{sec:magnus_params} shows how $n_{\text{groupElems}}$ is determined).
\texttt{GroupRange()} returns the flattened index range, and \texttt{Search()} (a modified binary search) returns the largest slice offset less than $i*n_{\text{groupElems}}$.
Each thread performs its own $O(\log(A.m))$ binary search\footnote{While a group- or subgroup-wide parallel search could reduce this cost, our approach performed well in practice.}. 
\texttt{ARange()} returns the range of elements in $\tilde{A}_\text{CSC}$ for a work group,
which calls \texttt{GetAEntry($groupRange.start,\dots$)} (not shown in the algorithm).
\texttt{GetAEntry($i,\dots$)} maps $i$ in flattened space to an element in $\tilde{A}_{\text{CSC}}$ as:
\begin{equation}
  \tilde{A}_{\text{CSC}}.\mathit{colPtr}[\mathit{slice}] + \frac{i - \mathit{sliceOffsets}[\mathit{slice}]}{\tilde{B}.\mathit{rowPtr}[\mathit{slice}+1] - \tilde{B}.\mathit{rowPtr}[\mathit{slice}]},
  \label{equ:get_A_entry}
\end{equation}
where $\mathit{slice} = \mathit{sliceRange}.\mathit{start}$ and $\mathit{slice} = \mathit{sliceRange}.\mathit{end}$ for the start and end elements, respectively, where $\dots$ in the function call denotes that the remaining arguments consist of the global variables and partition information.
The first term represents the starting column of $\tilde{A}_{\text{CSC}}$, and the second term gives the starting element within the column.
The numerator in the second term denotes the mapping of the flattened index $i$ to the local index within the current slice.
Dividing by the $\tilde{B}$ row size yields the element in the current column of $\tilde{A}_{\text{CSC}}$.

In step 2 (lines~\ref{alg:magnus_outer:step2_init}--\ref{alg:magnus_outer:step2_final}), updates to $\hat{C}.\mathit{rowPtr}$ determine the write locations for populating $\hat{C}$.
The objective is to write rows (or partial rows) of $\tilde{B}$ to contiguous locations in $\hat{C}$.
Each work group iterates over elements of $\tilde{A}_\text{CSC}$ in its partition and performs atomic updates to $\hat{C}.\mathit{rowPtr}$ using row sizes of $\tilde{B}$.
Because \texttt{AtomicAdd()} returns the previous value of the target address, we store these return values in local memory as the eventual write locations for populating $\hat{C}$.
\texttt{GetExcess()} detects partial $\tilde{B}$ rows and adjusts the atomic updates accordingly.
It returns the number of excess elements in the current $\tilde{B}$ row, that is, the elements that belong to one or more other work groups, using a calculation similar to \texttt{GetAEntry()}.
Note that we update $\hat{C}.rowPtr$, which was computed earlier in the setup phase.
Our implementation is in-place: the setup phase modifies elements $[2:\hat{C}.n+2)$, and the outer product modifies the shifted elements $[1:\hat{C}.n+1)$, yielding the correct $\hat{C}.rowPtr$ at completion of the outer product.

In step 3 (lines~\ref{alg:magnus_outer:step3_init}--\ref{alg:magnus_outer:step3_final}), we populate $\hat{C}$ by writing consecutive elements of $\tilde{B}$ to consecutive elements in $\hat{C}$.
These accesses are not fully coalesced because the rows of $\tilde{B}$ are not necessarily multiples of the cache line (Intel) or sector length (NVIDIA).
In practice, heavy rows usually consist of dense $\tilde{B}$ rows, enabling many coalesced writes and good performance for $g$MAGNUS.
The function \texttt{GetABEntries()} calls \texttt{GetAEntry()} as well as \texttt{GetBEntry()}, which is identical to \texttt{GetAEntry()} except it uses a modulo instead of a division in the second term of \autoref{equ:get_A_entry}.
When writing rows of $\tilde{B}$, we again adjust for partial rows.
Note that in both steps 2 and 3, a while loop adjusts the current slice, causing subgroup/warp divergence.
However, because the outer product is applied to heavy rows, each work group usually touches only a few slices, keeping the number of while-loop iterations low.

In summary, our static partitioning 
yields coalesced reads with high reuse for input matrices.
Writes to $\hat{C}$ are often highly coalesced as well since heavy rows tend to involve denser rows of $B$.
Unlike previous work that use dynamic global load balancing~\cite{blockReorg}, our local-only partitioning is lightweight, determined only by register-level calculations.
Additionally, since our algorithm is only applied to heavy rows, which means that $\tilde{A}_{\text{CSC}} \ll \hat{C}.nnz$, both the CSR-to-CSC conversion cost and the cost of the atomic operations in the first step of the outer product are often small compared to populating $\hat{C}$, as shown in~\autoref{sec:results}.

\subsection{The Multisplit Kernel}\label{sec:magnus_multisplit}


The construction of $\hat{C}_{\text{reord}}$ involves three kernels: a histogram kernel followed by a device-wide prefix sum computes $\hat{C}_{\text{reord}}.rowPtr$, and a multisplit kernel~\cite{gpu-multisplit} populates $\hat{C}_{\text{reord}}$.
We describe only the multisplit kernel, as the histogram kernel is a subset of its steps, namely local histogramming and updates to $\hat{C}_{\text{reord}}.rowPtr$.
Our design goal for multisplit is to divert the highly irregular permute phase to local memory, maximizing coalesced accesses to global memory.
\autoref{alg:multisplit} shows our multisplit kernel.
As in the outer product, $\bm{x_{\text{map}}}$ and a static partitioning of the flattened index space of $\hat{C}$ are used.
Compared to the outer product, the partitioning is simplified since we only need to index into $\hat{C}$ instead of both $\tilde{A}_{\text{CSC}}$ and $\tilde{B}$, removing the need for \texttt{GetEntry()} and \texttt{GetExcess()}.

Within a work group, multisplit has the following steps: local histogram, global update to $\hat{C}_{\text{reord}}.rowPtr$, local prefix sum, local permute, and global write to $\hat{C}_{\text{reord}}$.
There are four local memory arrays: $\mathit{offsetsLocal}_\ell$, $\mathit{offsetsGlobal}_\ell$, $\mathit{colBuff}_\ell$, and $\mathit{valBuff}_\ell$.
$\mathit{colBuff}_\ell$ and $\mathit{valBuff}_\ell$ store the local chunks, i.e., reordered column indices and values of $\hat{C}_{\text{reord}}$.
$\mathit{offsetsLocal}_\ell$ stores the local chunk offsets, where the elements of local chunk $i$ are stored at positions $[\mathit{offsetsLocal}_\ell[i]:\mathit{offsetsLocal}_\ell[i]+1)$ of $\mathit{colBuff}$ and $\mathit{valBuff}$.
$\mathit{offsetsGlobal}_\ell$ stores the global write positions: 
elements at positions $[\mathit{offsetsLocal}_\ell[i]:\mathit{offsetsLocal}_\ell[i]+1)$ of $\mathit{colBuff}$ and $\mathit{valBuff}$ are written to $\hat{C}_{\text{reord}}.col$ and $\hat{C}_{\text{reord}}.val$ starting at position $\mathit{offsetsGlobal}_\ell[i]$, respectively.

\begin{algorithm}[htbp]
    \setlength{\algomargin}{0pt}
    \algfontsize
    \SetCommentSty{emph}
    \DontPrintSemicolon
    \caption{\texttt{MultisplitKernel}}\label{alg:multisplit}
    \KwIn{$\bm{\hat{C}}$, $\bm{x_{\text{map}}}$,$n_{\text{chunks}}$, $n_{\text{groupElems}}$}
    \KwOut{$\bm{\hat{C}_{\text{reord}}}$}
    \KwLaunch{$\left\lceil \frac{\hat{C}.nnz}{n_{\text{workGroup}}}\right\rceil$, $n_{\text{workGroup}}$}
    \BlockComment{Local load balancing}\;
    $\mathit{groupRange} \gets \texttt{GroupRange}(n_{\text{groupElems}})$\;
    $rowRange \gets \texttt{Search}(\bm{\hat{C}.rowPtr}, \; \mathit{groupRange})$\;
    \BlockComment{Local histogram}\;
    $row \gets 0$, 
    $\mathit{offsetsLocal}_\ell \gets 0$\;\label{alg:multisplit:histo_init}
    \ForPar{$i \in [\mathit{groupRange.start}:\mathit{groupRange.end})$}{
        \While{$i \ge \bm{\hat{C}.rowPtr}[\mathit{rowRange.start}+row+1]$}{
            $row \gets row + 1$\;
        }
        $chunk \gets \texttt{GetChunk(}\bm{\hat{C}.col[i]},row, \dots\texttt{)}$\;
        \texttt{AtomicAdd($\mathit{offsetsLocal}_\ell[chunk+2],\; 1)$}\;
    }\label{alg:multisplit:histo_final}
    \texttt{SyncThreads()}\;

    \BlockComment{Updates to $\bm{\hat{C}_{\text{reord}}.chunkPtr}$}\;
    \ForPar{$i \in [0:n_{\text{groupRows}}\times n_{\text{chunks}})$}{\label{alg:multisplit:write_offsets_init}
        $\mathit{chunkElems} \gets \mathit{offsetsLocal}_\ell[chunk+2]$\;
            $chunk\gets \bm{x_{\text{map}}}[\mathit{rowRange.start}\times n_{\text{chunks}} + i]+1$\;
            $\mathit{offsetsGlobal}_\ell[i] \gets$ \texttt{AtomicAdd($\bm{\hat{C}_{\text{reord}}.\mathit{chunkPtr}}[\mathit{chunk}], \; \mathit{chunkElems}$}\texttt{)}\;
    }\label{alg:multisplit:write_offsets_final}
    \texttt{SyncThreads()}\;

    \BlockComment{Local offsets via prefix sum}\; \texttt{PrefixSumInPlace}($\mathit{offsetsLocal}_\ell[2:n_{\text{groupRows}}\times n_{\text{chunks}}+2)$\texttt{)}\;\label{alg:multisplit:pre_sum}
    \texttt{SyncThreads()}\;
    
    \BlockComment{Permute in local memory}\;
    $row \gets 0$\;
    \ForPar{$i \in [\mathit{groupRange.start}:\mathit{groupRange.end})$}{\label{alg:multisplit:reorder_init}
        \While{$i \ge \bm{\hat{C}.rowPtr}[\mathit{rowRange.start}+row+1]$}{
            $row \gets row + 1$\;
        }
        $chunk \gets \texttt{GetChunk(}\bm{\hat{C}.col[i]},row, \dots\texttt{)}$\;
        $dest \gets $\texttt{AtomicAdd($\mathit{offsetsLocal}_\ell[chunk+1],\; 1)$}\;
        $\mathit{colBuff}_\ell[dest] \gets \bm{\hat{C}.col}[i]$\;
        $\mathit{valBuff}_\ell[dest] \gets \bm{\hat{C}.val}[i]$\;
    }\label{alg:multisplit:reorder_final}
    \texttt{SyncThreads()}\;
    
    \BlockComment{Coalesced writes to global memory}\;
    $row \gets 0$\;
    \ForPar{$i \in [\mathit{groupRange.start}:\mathit{groupRange.end})$}{\label{alg:multisplit:glob_write_init}
        \While{$i \ge \bm{\hat{C}.rowPtr}[\mathit{rowRange.start}+row+1]$}{
            $row \gets row + 1$\;
        }
        $chunk \gets \texttt{GetChunk(}\mathit{colBuff}_\ell[i-\mathit{groupRange.start}], row, \dots\texttt{)}$\;
        $dest \gets \mathit{offsetsGlobal}_\ell[chunk] + i - \mathit{offsetsLocal}_\ell[chunk]$\;
        $\bm{\hat{C}_{\text{reord}}.col}[dest] \gets \mathit{colBuff}_\ell[i] - chunk\times \hat{C}_{\text{reord}}.m$\;
        $\bm{\hat{C}_{\text{reord}}.val}[dest] \gets \mathit{valBuff}_\ell[i]$\;
    }\label{alg:multisplit:glob_write_final}
\end{algorithm}

The size of $\mathit{colBuff}_\ell$ and $\mathit{valBuff}_\ell$ is $n_{\text{groupElems}}$.
The size of $\mathit{offsetsGlobal}_\ell$ and $\mathit{offsetsLocal}_\ell$ is $n_{\text{groupRows}}\times n_{\text{chunks}}$ and $n_{\text{groupRows}} \times n_{\text{chunks}}+1$, respectively.
The quantity
$n_{\text{groupRows}} = \left\lceil \frac{n_{\text{groupElems}} - 1}{\tau} \right\rceil + 1$
denotes the maximum number of rows covered by any work group.
The $n_{\text{groupRows}} \times n_{\text{chunks}}$ term accounts for row-boundary crossings, where we need to store a distinct set of $n_{\text{chunks}}$ offsets per row.
The additional 2 enables multiple in-place histogramming passes over $\mathit{offsetsLocal}_\ell$, first to count local chunks and then to update offsets during the local permute.

In the local histogram step (lines~\ref{alg:multisplit:histo_init}-\ref{alg:multisplit:histo_final}), the number of elements in each local chunk is counted.
Column indices of $\hat{C}$ are mapped to chunks using \texttt{GetChunk()}, which returns $chunk \gets row \times n_{\text{chunks}} + \left\lfloor \frac{\bm{\hat{C}.col}[i]}{\bm{\hat{C}_{\text{reord}}.m}}\right\rfloor$, where $row$ is the local row index in the current partition.
The local histogram is then used to update $\hat{C}_{\text{reord}}.rowPtr$
(lines~\ref{alg:multisplit:write_offsets_init}--\ref{alg:multisplit:write_offsets_final}).
As in the outer product kernel, we employ an in-place update scheme for $\hat{C}_{\text{reord}}.rowPtr$, where elements $[2, \hat{C}_{\text{reord}}.n+2)$ were updated in
the histogram kernel prior to the multisplit kernel.
Next, a work group-wide parallel prefix sum (\autoref{alg:multisplit:pre_sum}) of $\mathit{offsetsLocal}_\ell$ computes the local chunk offsets using vendor primitives, e.g., \texttt{cub::DeviceScan} on NVIDIA GPUs.
These offsets are used to then perform the local permute to get the local chunks (lines \ref{alg:multisplit:reorder_init}-\ref{alg:multisplit:reorder_final}).
The reordered column indices and values are stored in $\mathit{colBuff}_\ell$ and $\mathit{valBuff}_\ell$ at locations determined by atomically updating $\mathit{offsetsLocal}_\ell[i]$.
Finally, the local chunks are written to global memory, populating $\hat{C}_{\text{reord}}$ (lines~\ref{alg:multisplit:glob_write_init}-\ref{alg:multisplit:glob_write_final}).
To maximize coalesced writes, consecutive elements within a local chunk are written to consecutive elements in global memory:
elements $[\mathit{offsetsLocal}_\ell[i],\mathit{offsetsLocal}_\ell[i+1])$ of $\mathit{colBuff}_\ell$ and $\mathit{valBuff}_\ell$ are written to $\hat{C}_{\text{reord}}.col$ and $\hat{C}_{\text{reord}}.val$, respectively, starting at the locations stored in $\mathit{offsetsGlobal}_\ell[i]$.
This achieves our design goal of coalesced reads and writes, with irregular nonconsecutive writes diverted to local memory during the local permute phase.

Note that, as in the outer product, we do not achieve perfectly coalesced writes since the local chunk counts are not guaranteed to align with the cache line (or sector) size.
This behavior is expected given the irregularity of SpGEMM, yet the approach still yields substantially more coalesced writes than performing the reordering directly in global memory.
Our experiments show performance comparable to other multisplit implementations for more regular applications.
Additionally, although not shown in \autoref{alg:multisplit} for readability, we employ subgroup-private (warp-private) histograms, a common optimization that reduces atomic contention~\cite{mesman}.
In this scheme, $\mathit{offsetsLocal}_\ell$ has size $n_{\text{groupRows}}\times n_{\text{chunks}} \times n_{\text{subGroups}} + 1$, where subgroup $i$ updates elements $[n_{\text{groupRows}}\times n_{\text{chunks}}\times i, n_{\text{groupRows}}\times n_{\text{chunks}}\times (i+1))$.

Our algorithm has several additional benefits.
First, it performs few atomic operations relative to intermediate-element accesses: $O(\hat{C}_{\text{reord}}.n)$ atomics versus $O(\hat{C}_{\text{reord}}.nnz)$ reads and writes, where typically $\hat{C}_{\text{reord}}.nnz \gg \hat{C}_{\text{reord}}.n$ for heavy rows.
Atomic contention is also low, since threads within a work group update distinct elements of $\hat{C}_{\text{reord}}.rowPtr$, so at most $n_{\text{groupRows}}$ threads access the same element ($n_{\text{groupRows}} = 2$ in our experiments).
Second, we reduce register pressure to maximize occupancy.
In particular, we avoid nested loops, which are often unrolled by the compiler and increase register usage.
For example, in the final loop, even though the local intermediate elements are already ordered by chunk, we remap each element to its chunk number instead of using nested loops over chunks and elements.

\subsection{Parameters Selection}\label{sec:magnus_params}
In the initial host-side preprocessing step (\texttt{GetMagnusParams()} in \autoref{alg:magnus}), the $g$MAGNUS parameters are computed using inputs that are easily queried at runtime.
These inputs are $C.m$ (the number of columns of $C$), and the size in bytes of the data types used to store the CSR arrays.
We define $s_y$ as the number of bytes of the data type of variable $y$.
All quantities in this section are integers, so division is assumed to use floor rounding unless a ceiling is explicitly specified.
The computation of $\tau$, the large chunk threshold (i.e., the maximum capacity of the numeric-phase local memory hash map), is given by:
\begin{equation} 
\tau = \frac{s_{\text{LM}}}{\frac{1}{\alpha} \times s_{C.col}\times s_{C.val}}, 
\end{equation}
where $\alpha$ is the load factor used to reduce collisions.
Our implementation of hash accumulators uses $\alpha = 2$.
We also have the maximum dense accumulator size for numeric phase as $n_{\text{denseNumeric}} = s_{\text{LM}}/(s_{C.val}+s_{\text{bitMap}})$, which we use to calculate $n_{\text{chunksRequired}} = C.m/n_{\text{denseNumeric}}$, the minimum number of chunks per row required for local memory-only dense accumulation.
Rounding to the nearest power of two allows us to use bit shift operations instead of division when mapping column indices to chunks, as explained in \autoref{sec:magnus_multisplit}.

Before using $n_{\text{chunksRequired}}$ to compute $v_{\text{chunks}}$ (an array of size $n_{\text{levels}}$ containing the number of chunks per row per level), we need to calculate $n_{\text{chunksMax}}$, the maximum number of chunks that our multisplit kernel can process in a single pass (a single level).
Because the number of required chunks grows with $C.m$, $n_{\text{chunksMax}}$ represents the threshold at which $\mathit{offsetsLocal}_\ell$ and $\mathit{offsetsGlobal}_\ell$ no longer fit in local memory, beyond
which our multilevel algorithm is needed.
We compute $n_{\text{chunksMax}}$ by solving
\begin{equation}
\begin{split}
s_{\text{LM}} =\; &s_{\text{histo}} \times (n_{\text{groupRows}} \times n_{\text{subGroups}}\times      n_{\text{chunksMax}}+1)\; + \\
&s_{\hat{C}.rowPtr}\times n_{\text{groupRows}} \times n_{\text{chunksMax}}\; + \\ 
&\left(s_{C.col} + s_{C.val}\right)\times n_{\text{minGroupElems}}
\end{split}
\label{equ:multisplit_local_memory_bytes}
\end{equation}
for $n_{\text{chunksMax}}$, where the right-hand side denotes the total size in bytes of all local memory arrays.
The three terms represent the sizes in bytes of $\mathit{offsetsLocal}_\ell$, $\mathit{offsetsGlobal}_\ell$, and $\mathit{colBuff}_\ell$ and $\mathit{valBuff}_\ell$.
Multiplication by $n_{\text{subGroups}}$, the number of subgroups per work group, reflects histogram duplication across subgroups.
$n_{\text{minGroupElems}}$ denotes the minimum number of elements per work group in the flattened partition.
We set
$n_{\text{minGroupElems}} = \frac{s_{LM}}{2\times(s_{C.col} + s_{C.val})}$,
i.e., half of the available local memory is allocated to the locally permuted intermediate elements.
In practice, this choice balances local memory usage between the locally permuted elements and the histogram arrays, which we found yields high local memory utilization.
Solving gives us
\begin{equation}
n_{\text{chunksMax}} = \frac{s_{\text{LM}}-n_{\text{minGroupElems}} \times (s_{C.col} + s_{C.val})-s_{\text{histo}}}{n_{\text{groupRows}} \times \left( n_{\text{subGroups}}\times s_{\text{histo}}+ s_{\hat{C}.rowPtr}\right)}.
\end{equation}

With $n_{\text{chunksRequired}}$ and $n_{\text{chunksMax}}$, we can calculate the number of levels. 
If $n_{\text{chunksRequired}} < n_{\text{chunksMax}}$, we have one level with $n_{\text{chunksRequired}}$ chunks per row. Otherwise, we compute the number of levels by solving 
$n_{\text{chunksMax}}^{n_{\text{levels}}} = n_{\text{chunksRequired}}$.
Intuitively, this corresponds to the number of times we need to divide $C.m$ by $n_{\text{chunksMax}}$ to obtain our target local memory-only dense accumulator size.
This represents the hierarchical component of $g$MAGNUS, giving us
\begin{equation} 
n_{\text{levels}} = \left \lceil \frac{\log_2(n_{\text{chunksRequired}})}{\log_2(n_{\text{chunksMax}})}\right\rceil. 
\end{equation} 
Since this gives us the minimum number of levels, one level may have fewer chunks than $n_{\text{chunksMax}}$. 
We set this as the last level: $v_{\text{chunks}}[i] = n_{\text{chunksMax}}$ for $i \in [0,n_{\text{levels}}-1)$ and $v_{\text{chunks}}[n_{\text{levels}}-1] = n_{\text{chunksRequired}} - n_{\text{chunksMax}}^{n_{\text{levels}}-1}$. 
Using more chunks in earlier levels acts as a filter: subdividing the row into smaller parts produces fewer heavy chunks, increasing the chance of later levels reordering fewer elements.
An additional optimization is that we floor $n_{\text{chunksRequired}}$ and $n_{\text{chunksMax}}$ to the nearest power of two.
This lets us map column indices to chunks using lower-latency bit shifts instead of integer division.

The final key parameter is the number of elements per work group in the flattened space, $n_{\text{groupElems}}$, which is kernel-dependent.
In general, we choose this quantity as large as possible to maximize either local memory utilization in the multisplit kernel or input reuse in the outer product kernel.
For the multisplit kernel, $n_{\text{groupElems}}$ is the number of elements that fit in the available shared memory after accounting for the local histograms, rounded down to the nearest multiple of the work-group size.
For the outer product kernel, we use
$n_{\text{groupElems}} = \frac{s_{\text{LM}}}{s_{\hat{C}.rowPtr}}$,
which is the maximum size of $\mathit{offsetsLocal}_\ell$.

\subsection{Accumulation}\label{sec:accum}
$g$MAGNUS is agnostic to the specific local-memory accumulator: any optimized local-memory accumulator can be used for light rows and chunks, and any optimized local-memory dense accumulator can be used for heavy chunks.
While we use simplified implementations of both, integrating more optimized library accumulators into $g$MAGNUS could further improve performance and is left for future work.
Our implementation uses a 
variant of the common \emph{binning} approach~\cite{OpSparse}, in which lighter rows are assigned smaller work-group sizes according to their intermediate product sizes.
This requires a setup phase that uses a kernel similar to multisplit to map rows and chunks to bins, where each bin corresponds to a work-group size.
We then launch one asynchronous kernel per bin, with one work group per row or chunk within the bin.
For light rows, we use a standard hash map with modulo hashing and linear probing, followed by a bitonic sort to produce the final row of $C$ (or subrow, for light chunks).
For heavy chunks in the numeric phase, we use dense accumulation to merge elements, followed by a work-group-wide prefix sum over the bitmap to generate the sorted final chunk.
The index arrays returned by \texttt{FlaggedPartition()} provide the mapping back to the original row and chunk orderings (the inverse maps of those in \autoref{alg:magnus_outer} and \autoref{alg:multisplit}), which we use when writing the final rows of $C$.

\section{Experimental Results}\label{sec:results}

\subsection{Experimental Setup}
We evaluate our SYCL and CUDA implementations against five state-of-the-art SpGEMM implementations spanning vendor libraries
and prior academic work: Intel MKL~\cite{mkl}, Kokkos~\cite{kokkos,kokkos2}, NVIDIA cuSPARSE~\cite{cuSPARSE}, TileSpGEMM~\cite{tileSpGEMM}, and OpSparse~\cite{OpSparse}.
All methods are compiled with CUDA 13.1.1 or oneAPI 2025.3.0.
TileSpGEMM and OpSparse are selected as academic baselines because they are recent, algorithmically distinct, and have been evaluated against a broad set of prior approaches.
For cuSPARSE, we test all three available algorithms, including five separate runs of algorithm 3, with parameters 0.1, 0.2, 0.3, 0.4, and 0.5.
This parameter controls the fraction of intermediate products processed at once as a way to reduce memory consumption.
We report the minimum execution time across all seven cuSPARSE configurations, where algorithm 1 is usually the fastest followed by 2 and then 3.
For all algorithms, we report the total end-to-end time, which is the sum of any pre-processing and/or setup, compute (symbolic and numeric), and post-processing phases.
For all experiments, we perform one warmup run followed by 20 timed runs and report the mean time.
Some baselines failed to run to completion on certain matrices due to segmentation faults, illegal memory accesses, out-of-memory errors, or timeouts (using a cutoff of 20 minutes).
\autoref{tab:gpu_specs} shows our test data center GPUs: Intel Ponte Vecchio (PVC) 1100 and NVIDIA H200.
The SYCL implementations (MKL and Kokkos) are evaluated on the Intel PVC GPU, and the CUDA implementations are evaluated on H200 (cuSPARSE, Kokkos, TileSpGEMM, and OpSparse).

\begin{table}[htbp]
\caption{Specifications of the test GPUs.}
\vspace*{-5pt}
\begin{center}
\resizebox{\columnwidth}{!}{
\begin{tabular}{lrr}
\hline
Architecture & Intel PVC 1100 & NVIDIA H200 \\
\hline
Memory & 48 GB & 141 GB \\
Peak memory bandwidth & $\sim$1.2 TB/s & $\sim$4.8 TB/s \\
Max local memory per Xe-core / SM & 128 KB & 228 KB \\
L1 / shared cache per Xe-core / SM & 512 KB & 256 KB \\
L2 cache (global) & 408 MB & 50 MB \\
\hline
\end{tabular}
}
\label{tab:gpu_specs}
\end{center}
\end{table}

We evaluate $g$MAGNUS on two matrix data sets: the SuiteSparse matrix collection~\cite{suitesparse} and recursive model power-law matrices (RMats)~\cite{rmat}.
For SuiteSparse, we compute $A^2$, which is standard practice in SpGEMM evaluation.
Following prior work~\cite{HSMU-SpGEMM,tileSpGEMM}, we include all matrices (330 in total) that require at least 100 million floating-point operations and run to completion for $g$MAGNUS and at least one baseline on at least one GPU.
We also highlight a subset of 20 matrices (shown in \autoref{tab:select_suitesparse_specs}) with the largest $C.nnz$ that satisfy two additional criteria: they run to completion on both GPUs, and at least $10\%$ of rows are classified as heavy on both GPUs.

For the RMat matrices, we use PaRMAT~\cite{PaRMAT} with the standard Graph500 parameters ($a = 0.57$, $b = c = 0.19$) to generate various pairs of $A$ and $B$. The matrix \emph{scale} ranges from 15 to 20, corresponding to matrices with $2^{scale}$ rows and columns.
The density of $A$ varies as $A.nnz/A.n \in \{4, 8, 12, \ldots, 64\}$, while the density of $B$ varies as $B.nnz/B.n \in \{4, 8, 12, 16\}$.
Increasing the density of either input matrix raises the memory cost of storing $C$ and reduces the largest matrix scale that can be evaluated.
When choosing which input density to vary more aggressively, we increase $A.nnz/A.n$ rather than $B.nnz/B.n$.
This increases the number of distinct $B$ rows accessed rather than the length of individual $B$ rows, introducing additional indirection and more irregular memory accesses.
For each scale, we evaluate all combinations of $A$ and $B$ for a total of 384 matrices.

These RMat matrices are challenging because their intermediate products vary widely in size, the distribution of column indices spans the full column range of $C$, and their irregular structure limits the effectiveness of accumulators that exploit structural matrix locality. \autoref{tab:rmat_nnz4_specs} shows a representative set of RMat matrices.
For this set, $B.nnz/B.n = 8$ and $A.nnz/A.n \in \{4, 8, 16, 32\}$.
Due to the nonuniform 
structure that arises from the Graph500 parameters, the ratio ${\hat{C}}.{nnz}/C.nnz$ increases as $A.nnz/A.n$ increases.
This means that as $A.nnz/A.n$ increases, the intermediate matrix grows faster than the final matrix, increasing the amount of accumulation work required per nonzero in the final matrix.

\begin{table}[tbp]
\caption{Properties of the representative SuiteSparse matrices.}
\vspace*{-5pt}
\begin{center}
\resizebox{\tabwidth}{!}{
\begin{tabular}{lrrrrr}
\hline
Matrix & $A.n$ & $A.nnz$ & $\frac{A.nnz}{A.n}$ & $A^2.nnz$ & $\frac{A^2.nnz}{A.n}$\\
\hline
para-9 & 155,924 & 5,416,358 & 34.7 & 78,088,896 & 500.8 \\
Stanford\_Berkeley & 683,446 & 7,583,376 & 11.1 & 78,130,972 & 114.3 \\
soc-Slashdot0902 & 82,168 & 948,464 & 11.5 & 81,362,487 & 990.2 \\
HTC\_336\_4438 & 226,340 & 904,522 & 4.0 & 83,300,312 & 368.0 \\
bloweya & 30,004 & 150,009 & 5.0 & 100,360,010 & 3344.9 \\
a0nsdsil & 80,016 & 355,034 & 4.4 & 175,955,042 & 2199.0 \\
c-57 & 37,833 & 405,197 & 10.7 & 177,766,591 & 4698.7 \\
brainpc2 & 27,607 & 179,395 & 6.5 & 190,743,619 & 6909.2 \\
TSOPF\_FS\_b39\_c7 & 28,216 & 730,080 & 25.9 & 199,299,980 & 7063.4 \\
hangGlider\_5 & 16,011 & 155,246 & 9.7 & 202,569,269 & 12651.9 \\
in-2004 & 1,382,908 & 16,917,053 & 12.2 & 213,255,458 & 154.2 \\
net150 & 43,520 & 3,121,200 & 71.7 & 238,012,852 & 5469.0 \\
eu-2005 & 862,664 & 19,235,140 & 22.3 & 284,177,131 & 329.4 \\
vsp\_south31\_slptsk & 39,668 & 379,828 & 9.6 & 389,540,274 & 9820.0 \\
vsp\_model1\_crew1\_cr42\_south31 & 45,101 & 379,952 & 8.4 & 394,768,783 & 8753.0 \\
c-big & 345,241 & 2,341,011 & 6.8 & 447,991,461 & 1297.6 \\
pkustk12 & 94,653 & 7,512,317 & 79.4 & 474,804,911 & 5016.3 \\
mult\_dcop\_02 & 25,187 & 193,276 & 7.7 & 518,559,249 & 20588.4 \\
wb-edu & 9,845,725 & 57,156,537 & 5.8 & 630,077,764 & 64.0 \\
rajat28 & 87,190 & 607,235 & 7.0 & 898,546,696 & 10305.6 \\
\hline
\end{tabular}
}
\label{tab:select_suitesparse_specs}
\end{center}
\end{table}

\begin{table}[tbp]
\caption{Properties of the representative RMat matrices.
For this set, $B.nnz/B.n$ is fixed at 8.}
\vspace*{-5pt}
\begin{center}
\resizebox{\tabwidth}{!}{
\begin{tabular}{c|cc|cc|cc|cc}
\hline
\multirow{3}{*}{\begin{tabular}{c}Matrix\\Scale\end{tabular}} & \multicolumn{8}{c}{$A.nnz/A.n$} \\
\cline{2-9}
& \multicolumn{2}{c|}{4} & \multicolumn{2}{c|}{8} & \multicolumn{2}{c|}{16} & \multicolumn{2}{c}{32} \\
\cline{2-9}
& $C.nnz/C.n$ & ${\hat{C}}.{nnz}/\hat{C}.n$ & $C.nnz/C.n$ & ${\hat{C}}.{nnz}/\hat{C}.n$ & $C.nnz/C.n$ & ${\hat{C}}.{nnz}/\hat{C}.n$ & $C.nnz/C.n$ & ${\hat{C}}.{nnz}/\hat{C}.n$ \\
\hline
15 & 477.6 & 819.9 & 782.3 & 1548.9 & 1233.3 & 2841.0 & 1895.6 & 5149.9 \\
16 & 633.5 & 1068.2 & 1048.4 & 2021.2 & 1679.4 & 3746.7 & 2597.7 & 6779.1 \\
17 & 835.0 & 1382.7 & 1397.9 & 2628.5 & 2267.3 & 4911.0 & 3574.5 & 8979.8 \\
18 & 1112.7 & 1802.7 & 1875.6 & 3432.7 & 3075.3 & 6427.6 & 4896.9 & 11805.3 \\
19 & 1471.9 & 2342.3 & 2496.0 & 4472.1 & N/A & N/A & N/A & N/A \\
20 & 1944.1 & 3043.1 & N/A & N/A & N/A & N/A & N/A & N/A \\
\hline
\end{tabular}
}
\label{tab:rmat_nnz4_specs}
\end{center}
\end{table}
\subsection{SpGEMM Evaluation}\label{sec:results_spgemm}

\autoref{tab:performance_summary} shows the geometric-mean (geomean) speedup of $g$MAGNUS over the five baselines across all datasets and GPUs.
The results demonstrate that $g$MAGNUS is faster than all baselines, with a geomean speedup ranging from $1.19\times$ over OpSparse to $5.52\times$ over TileSpGEMM.
Failed baseline runs are excluded from the geomean calculation.
On H200, $g$MAGNUS and cuSPARSE ran to completion for all 330 matrices, whereas Kokkos, TileSpGEMM, and OpSparse completed 311, 271, and 295 matrices, respectively.
On PVC, $g$MAGNUS and MKL ran to completion for 312 matrices (the remaining 18 excluded because $C$ did not fit in device memory), whereas Kokkos completed 295 matrices.

\begin{figure*}[!htbp]
\centering
\setlength{\tabcolsep}{0pt}
\begin{tabular}{@{}c@{\hspace{-2em}}c@{}}
\raisebox{10pt}{\rotatebox[origin=c]{90}{$g$MAGNUS Speedup}} &
\begin{minipage}{\linewidth}
\centering
    \includegraphics[width=\wideBarScale\linewidth]{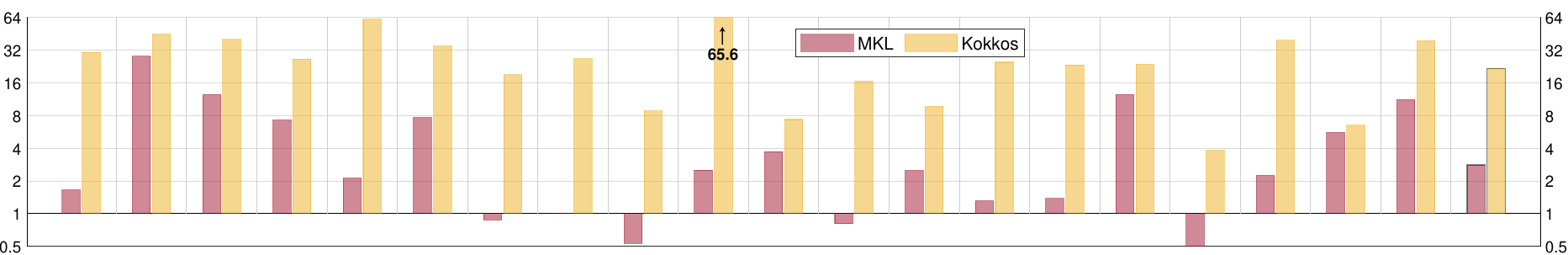} \\
    \includegraphics[width=\wideBarScale\linewidth]{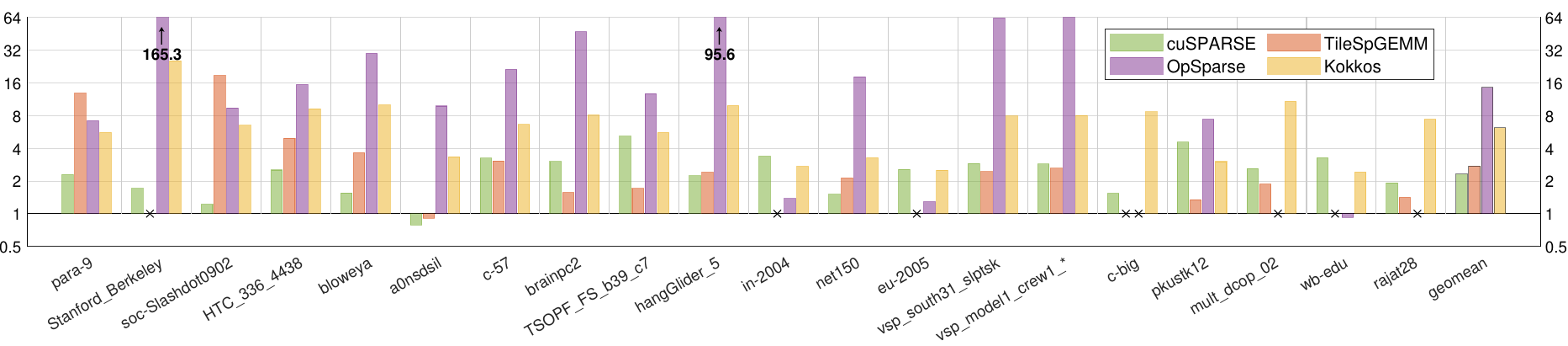}
\end{minipage}
\end{tabular}
\caption{Speedup in log scale for the 20 representative SuiteSparse matrices on Intel PVC (top) and NVIDIA H200 (bottom).  The geomean is shown in the last group of bars.  The $\times$-shaped markers denote failed runs.}
\label{fig:baselines_suitesparse}
\end{figure*}

\autoref{fig:baselines_suitesparse} shows the $g$MAGNUS speedup (ratio of the baseline time to $g$MAGNUS time) in log scale for the 20 representative matrices from \autoref{tab:select_suitesparse_specs}.
$g$MAGNUS is fastest for 34 out of 40 instances (20 matrices and 2 GPUs), with a geomean speedup of $2.81\times$, $13.98\times$, $2.32\times$, $2.74\times$, and $14.51\times$ over MKL, Kokkos, cuSPARSE, TileSpGEMM, and OpSparse, respectively.
Besides TileSpGEMM, the geomean speedup across the representative set increased compared to the the full 330 matrices, highlighting the benefit of $g$MAGNUS on large matrices with many heavy rows.
The speedup calculation excludes timeouts, for which TileSpGEMM had the highest rate, failing often for larger matrices.
The vendor libraries were the most competitive, likely due to adaptive strategies that provide robust performance across a wide range of applications.
OpSparse and Kokkos were the least competitive, demonstrating the drawbacks of global hash-map accumulators.
This is most apparent for highly irregular applications, such as Stanford\_Berkeley (web graph) and the vsp matrices (random unweighted graphs), and for applications where the input nonzeros are distributed broadly across the column range, such as hangGlider\_5.

\autoref{tab:select_suitesparse_magnus_params} shows runtime parameters of $g$MAGNUS.
The first column shows the total size, in GB, of the CSR arrays that compose $\hat{C}$, which ranges from 0.47 to 21.41.
The H200 values are always lower because its larger local memory results in fewer heavy rows.
Columns 2-3 show the fraction of intermediate product elements from heavy rows and the fraction of heavy rows, demonstrating that a small number of rows can produce large intermediate products.
For example, in Stanford\_Berkeley, less than 1\% of heavy rows account for half of the total intermediate product elements.
The last column shows the required number of chunks, notated as
\emph{required number of chunks: level 0 number of chunks, level 1 number of chunks, $\dots$}. The required number of chunks varies significantly due to the wide range of matrix sizes, with 7 of these matrices requiring more than one level and one matrix requiring 3 levels (wb-edu on PVC). In general, PVC requires twice as many chunks due to its smaller local memory.

\begin{table}[tbp]
\caption{Geometric-mean speedup of $g$MAGNUS over five baselines.}
\centering
\resizebox{\tabwidth}{!}{
\begin{tabular}{lcccccc}
\hline
& MKL & \multicolumn{2}{c}{Kokkos} & cuSPARSE & TileSpGEMM & OpSparse \\
& PVC & PVC & H200 & H200 & H200 & H200 \\
\hline
SuiteSparse & 1.28 & 4.92 & 2.66 & 1.74 & 5.52 & 1.19 \\
RMat & 3.33 & 26.78 & 8.53 & 1.91 & 11.43 & 7.94 \\
\hline
\textbf{Total} & \textbf{1.98} & \multicolumn{2}{c}{\textbf{6.63}} & \textbf{1.81} & \textbf{7.62} & \textbf{3.29} \\
\hline
\end{tabular}
}
\label{tab:performance_summary}
\end{table}

\begin{figure*}[!htbp]
\def\scaleTop{0.22}
\def\scaleBot{0.22}
\centering
\setlength{\tabcolsep}{2pt}
\begin{tabular}{@{}c@{\hspace{-2em}}c@{}}
\raisebox{10pt}{\rotatebox[origin=c]{90}{$g$MAGNUS Speedup}} &
\begin{minipage}{\linewidth}
\centering
\begin{tabular}{cccc}
$\boldsymbol{A.nnz/A.n=4}$ & $\boldsymbol{A.nnz/A.n=8}$ & $\boldsymbol{A.nnz/A.n=16}$ & $\boldsymbol{A.nnz/A.n=32}$ \\
\includegraphics[width=\scaleTop\linewidth]{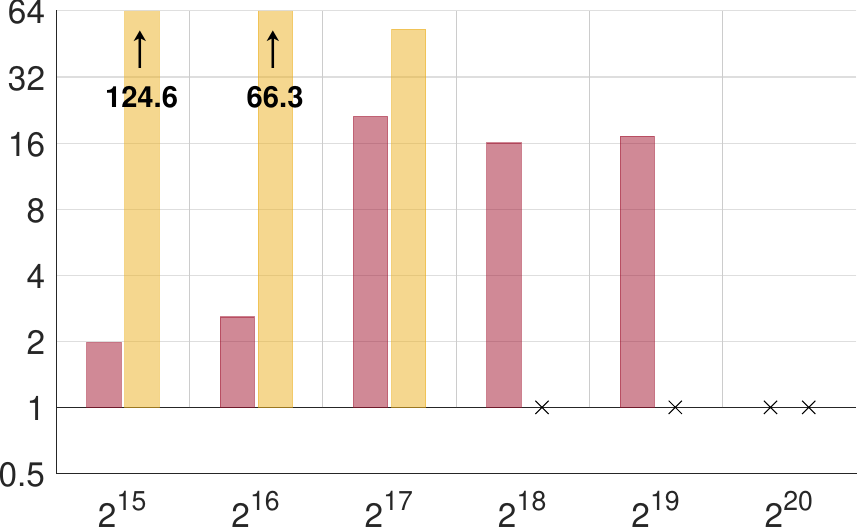} &
\includegraphics[width=\scaleTop\linewidth]
{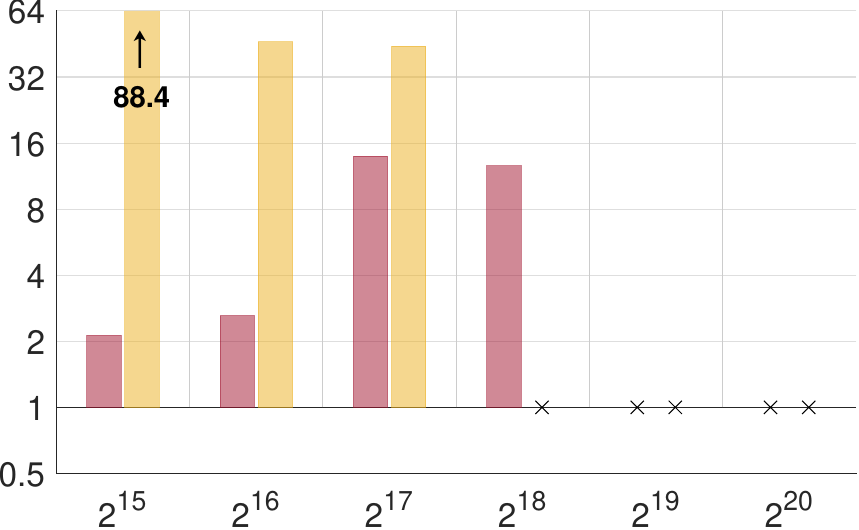} &
\includegraphics[width=\scaleTop\linewidth]
{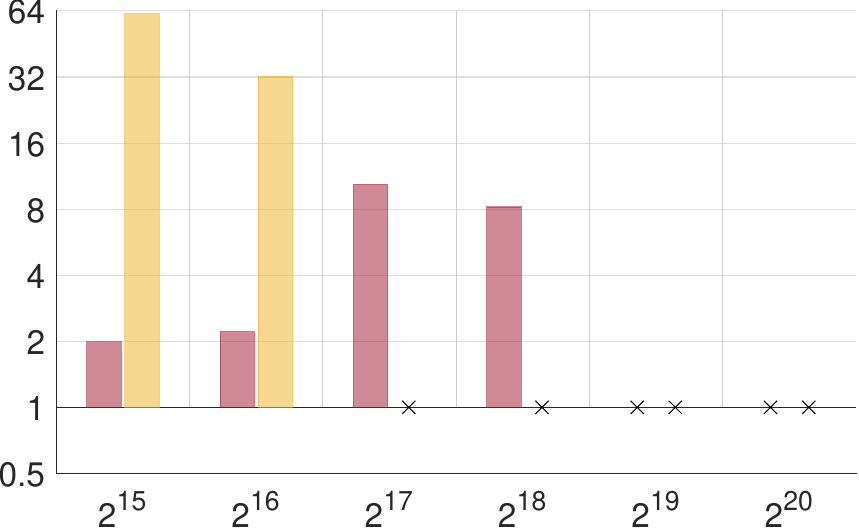} &
\includegraphics[width=\scaleTop\linewidth]{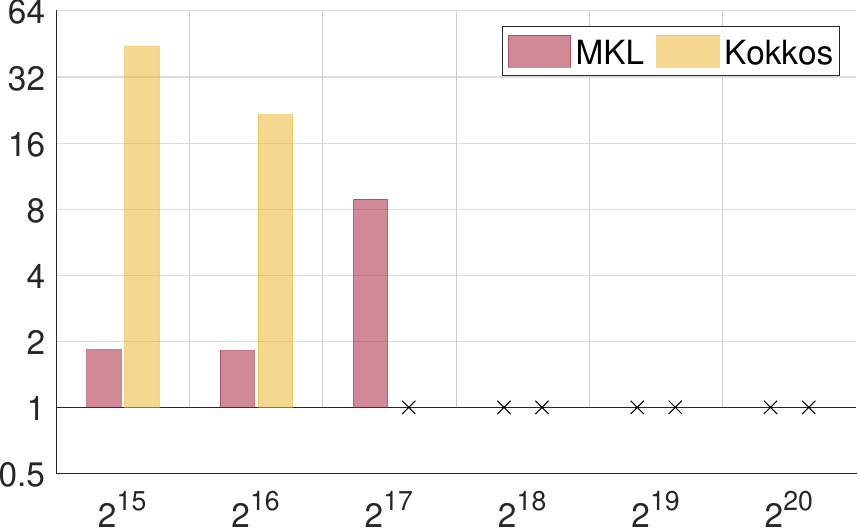} \\
\includegraphics[width=\scaleBot\linewidth]{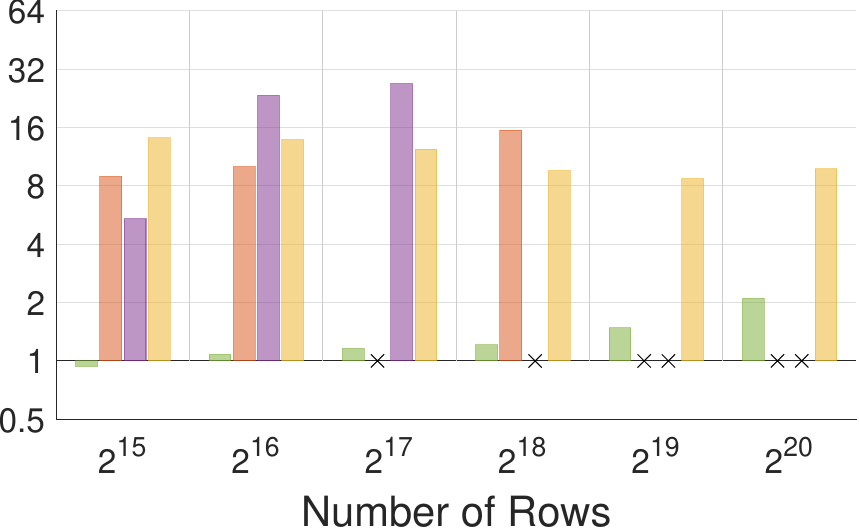} &
\includegraphics[width=\scaleBot\linewidth]
{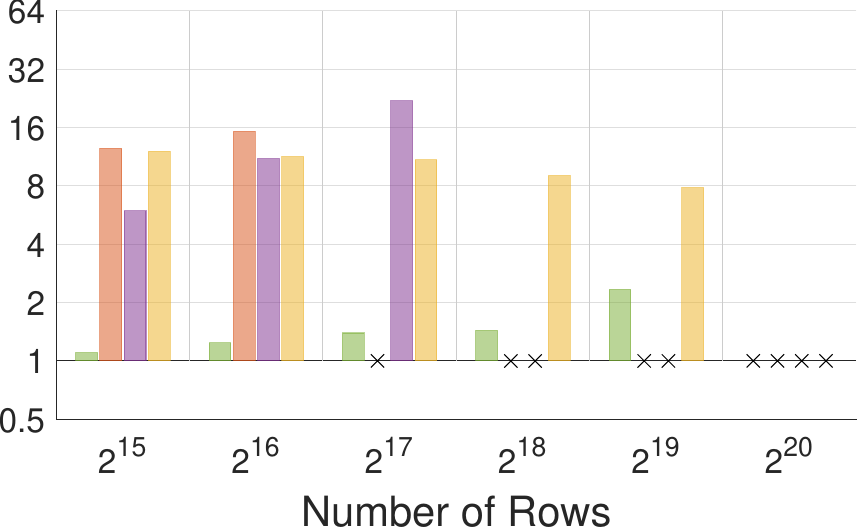} &
\includegraphics[width=\scaleBot\linewidth]
{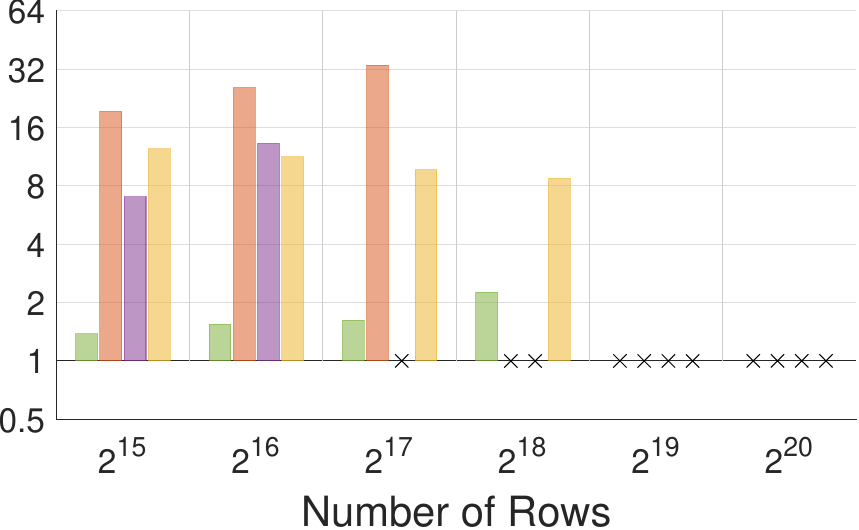} &
\includegraphics[width=\scaleBot\linewidth]{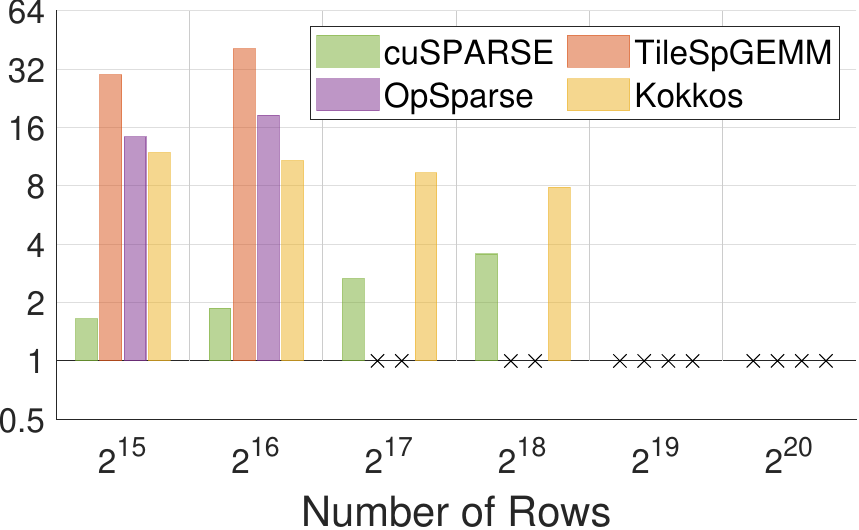} \\
\end{tabular}
\end{minipage}
\end{tabular}
\caption{Speedup in log scale vs number of rows of $C$ for the representative matrices from the RMat matrix set on Intel PVC (top row) and NVIDIA H200 (bottom row).  From left to right, the figures correspond to increasing $A.nnz/A.n$ (nonzeros per row of $A$), where $\times$ markers denotes failed runs.}
\label{fig:rmat_baselines}
\end{figure*}

\begin{table}[tbp]
\caption{$g$MAGNUS runtime parameters for the 20 representative SuiteSparse matrices.
\emph{Heavy IP \%}: fraction of total intermediate product elements from heavy rows.
\emph{Heavy Row \%}: fraction of heavy rows.
\emph{Num. Chunks}: 
number of chunks per row.
}
\vspace*{-3pt}
\begin{center}
\resizebox{\tabwidth}{!}{
\begin{tabular}{l|cc|cc|cc|cc}
\hline
\multirow{2}{*}{} & \multicolumn{2}{c|}{$\hat{C}$  Size (GB)} & \multicolumn{2}{c|}{Heavy IP \%} & \multicolumn{2}{c|}{Heavy Row \%} & \multicolumn{2}{c}{Num. Chunks} \\
\cline{2-9}
& PVC & H200 & PVC & H200 & PVC & H200 & PVC & H200 \\
\hline
para-9 & 0.45 & 0.44 & 0.21 & 0.21 & 0.0457 & 0.0445 & 64: 32, 2 & 32 \\
Stanford\_Berkeley & 0.88 & 0.74 & 0.50 & 0.42 & 0.0038 & 0.0023 & 256: 32, 8 & 128: 64, 2 \\ 
soc-Slashdot0902 & 0.63 & 0.47 & 0.67 & 0.51 & 0.0894 & 0.0470 & 32 & 16 \\ 
HTC\_336\_4438 & 2.13 & 2.13 & 0.98 & 0.98 & 0.0550 & 0.0550 & 64: 32, 2 & 32 \\ 
bloweya & 0.80 & 0.80 & 0.99 & 0.99 & 0.3334 & 0.3334 & 8 & 4 \\ 
a0nsdsil & 1.61 & 0.40 & 0.99 & 0.25 & 0.4375 & 0.0627 & 32 & 16 \\ c-57 & 5.51 & 5.45 & 0.99 & 0.98 & 0.4372 & 0.4053 & 16 & 8 \\ 
brainpc2 & 3.81 & 3.81 & 0.99 & 0.99 & 0.4999 & 0.4999 & 8 & 4 \\ 
TSOPF\_FS\_b39\_c7 & 15.93 & 15.93 & 0.99 & 0.99 & 0.4986 & 0.4986 & 8 & 4 \\ 
hangGlider\_5 & 1.63 & 1.63 & 0.99 & 0.99 & 0.8888 & 0.8888 & 4 & 2 \\ 
in-2004 & 4.53 & 4.43 & 0.33 & 0.32 & 0.0068 & 0.0051 & 512: 32, 16 & 256: 64, 4 \\ 
net150 & 3.65 & 3.65 & 0.98 & 0.98 & 0.6273 & 0.6273 & 16 & 8 \\ 
eu-2005 & 2.18 & 1.06 & 0.32 & 0.16 & 0.0381 & 0.0069 & 256: 32, 8 & 128: 64, 2 \\ 
vsp\_south31\_* & 5.38 & 5.33 & 0.99 & 0.98 & 0.5802 & 0.5533 & 16 & 8 \\
vsp\_model1\_* & 5.15 & 5.12 & 0.96 & 0.96 & 0.4430 & 0.4294 & 16 & 8 \\ 
c-big & 3.50 & 3.12 & 0.92 & 0.82 & 0.0766 & 0.0567 & 128: 32, 4 & 64 \\ 
pkustk12 & 21.41 & 19.67 & 0.99 & 0.92 & 0.9937 & 0.9062 & 32 & 16 \\ 
mult\_dcop\_02 & 4.16 & 4.16 & 0.99 & 0.99 & 0.9042 & 0.9042 & 8 & 4 \\ 
wb-edu & 2.03 & 1.66 & 0.16 & 0.13 & 0.0013 & 0.0004 & 4096: 32, 32, 4 & 2048: 64, 32 \\ 
rajat28 & 7.19 & 6.88 & 0.99 & 0.94 & 0.4052 & 0.3358 & 32 & 16 \\
\hline
\end{tabular}
}
\label{tab:select_suitesparse_magnus_params}
\end{center}
\end{table}

\begin{table}[tbp]
\caption{$g$MAGNUS runtime parameters for a representative set of RMat matrices, 
with $A.nnz/A.n = 4$ and $B.nnz/B.n = 8$.
\emph{Heavy IP \%}: fraction of total intermediate product elements from heavy rows. 
\emph{Heavy Row \%}: fraction of heavy rows.
\emph{Num. Chunks}: 
number of chunks per row.}
\vspace*{-3pt}
\begin{center}
\resizebox{\tabwidth}{!}{
\begin{tabular}{c|cc|cc|cc|cc}
\hline
\multirow{2}{*}{\begin{tabular}{c}Matrix\\Scale\end{tabular}} & \multicolumn{2}{c|}{$\hat{C}$ Size (GB)} & \multicolumn{2}{c|}{Heavy IP \%} & \multicolumn{2}{c|}{Heavy Row \%} & \multicolumn{2}{c}{Num. Chunks} \\
\cline{2-9}
& PVC & H200 & PVC & H200 & PVC & H200 & PVC & H200 \\
\hline
15 & 0.14 & 0.10 & 0.64 & 0.47 & 0.0487 & 0.0227 & 8 & 4 \\
16 & 0.41 & 0.34 & 0.73 & 0.60 & 0.0599 & 0.0345 & 16 & 8 \\
17 & 1.17 & 0.99 & 0.80 & 0.69 & 0.0718 & 0.0424 & 32 & 16 \\
18 & 3.24 & 2.91 & 0.86 & 0.77 & 0.0826 & 0.0538 & 64: 32, 2 & 32 \\
19 & 8.87 & 8.10 & 0.90 & 0.82 & 0.0941 & 0.0613 & 128: 32, 4 & 64 \\
20 & N/A & 22.60 & N/A & 0.89 & N/A & 0.0745 & N/A & 128: 64, 2 \\
\hline
\end{tabular}
}
\label{tab:rmat_nnz4_magnus_params}
\end{center}
\end{table}
For the RMat matrices, $g$MAGNUS demonstrates higher speedups than for SuiteSparse, with a geomean speedup ranging from $1.91\times$ over cuSPARSE to $26.78\times$ over Kokkos, as shown in \autoref{tab:performance_summary}.
\autoref{tab:rmat_nnz4_magnus_params} shows the runtime parameters used by $g$MAGNUS for $A.nnz/A.n = 4$ and $B.nnz/B.n = 8$. 
As expected, the required number of chunks is tied solely to matrix scale, doubling with each scale increment. 
When the maximum number of chunks is reached (scale 18 for PVC, scale 20 for H200), multiple levels are employed. 
Additionally, the fraction of heavy rows and heavy intermediate products increases with matrix scale, even though the densities of $A$ and $B$ are fixed, which further underscores the challenges of scaling SpGEMM to massive RMats.

\autoref{fig:rmat_baselines} shows the $g$MAGNUS speedup versus matrix scale for $A.nnz/A.n = \{4, 8, 16, 32\}$, which shows important performance trends as both matrix scale and $A.nnz/A.n$ increase.
On PVC, Kokkos times out for larger matrix scales and is orders of magnitude slower when it completes. 
While MKL is competitive with $g$MAGNUS at smaller matrix scales, it becomes orders of magnitude slower beyond scale 16. 
On H200, all baselines except cuSPARSE are orders of magnitude slower than $g$MAGNUS, with TileSpGEMM and OpSparse failing for larger scales. 
cuSPARSE remains competitive for smaller sizes and lower values of $A.nnz/A.n$, but diverges from $g$MAGNUS as these parameters increase. 
Overall, $g$MAGNUS exhibits the best scaling as the dimensions and density scale up.
\subsection{Performance of $g$MAGNUS Core Kernels}


\begin{figure*}[!htbp]
\centering
\includegraphics[width=\wideBarScale\linewidth]{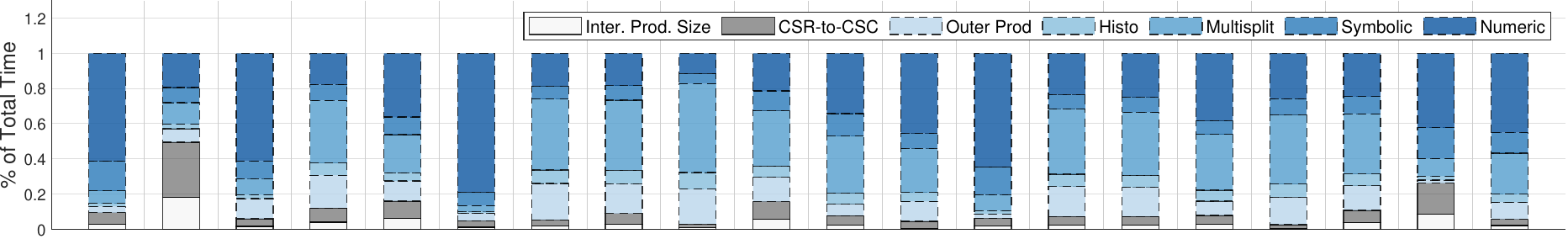}
\vspace*{-1pt}
\caption{Performance breakdown of $g$MAGNUS phases for the representative SuiteSparse matrices on H200.}
\label{fig:phases_suitesparse_H200}
\end{figure*}

\begin{figure*}[!htbp]
\centering
\includegraphics[width=\wideBarScale\linewidth]{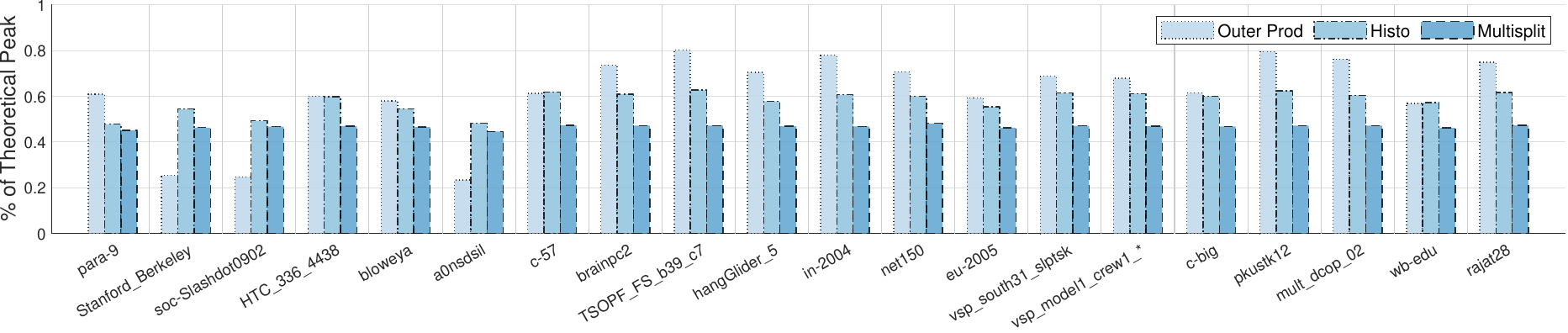}
\vspace*{-1pt}
\caption{Comparison of the core kernels in $g$MAGNUS to their theoretical upper bound for the representative SuiteSparse matrices on H200.}
\label{fig:sol_suitesparse_H200}
\end{figure*}


We evaluate the core kernels of $g$MAGNUS against their ``speed of light'', i.e., their theoretical peak performance.
The theoretical peak time of a kernel is its theoretical minimum data volume divided by the peak memory bandwidth, where the peak bandwidth is measured using simple memory copy benchmarks (e.g., \texttt{bandwidthTest} included in CUDA).
For example, the theoretical minimum data volume of the histogram kernel (which computes $\hat{C}_{\text{reord}}.rowPtr$) is: 
\begin{equation} 
\hat{C}.nnz\times s_{\hat{C}.col} + (\hat{C}.n+1)\times s_{\hat{C}.rowPtr} + \hat{C}_{\text{reord}}.n\times s_{\hat{C}_{\text{reord}}.rowPtr}.
\end{equation}
The first term represents reading $\hat{C}$, the second term represents reading $\hat{C}.rowPtr$, and the third term represents writing to $\hat{C}_{\text{reord}}.rowPtr$.
We only present results for H200, since PVC shows the same 
trends.

\autoref{fig:phases_suitesparse_H200} shows the breakdown of $g$MAGNUS into its performance-critical phases, and \autoref{fig:sol_suitesparse_H200} compares the core kernels against their speed-of-light performance.
Because the fraction of heavy intermediate products varies across matrices, the time spent in each SpGEMM phase also varies widely.
For example, para-9, soc-Slashdot0902, a0nsdsil, eu-2005, wb-edu, and rajat28 spend most of their time in the symbolic and numeric phases because they have the smallest fraction of heavy intermediate products, meaning that most intermediate products belong to light rows.
For the remaining matrices, heavy-row processing accounts for a significant portion of runtime and often exceeds half of the total execution time.
The outer product kernel achieves more than $50\%$ of theoretical peak performance for most matrices, and most of the largest matrices reach $60$--$80\%$ of peak.
The cases below $50\%$ typically arise when dense columns of $A_{\text{CSC}}$ are multiplied by highly sparse rows of $\tilde{B}$.
This structure reduces coalesced writes to $\hat{C}$ and increases atomic updates to $\hat{C}.rowPtr$, since the number of atomic operations is proportional to $A_{\text{CSC}}.nnz$.

For the histogram kernel, we achieve approximately $40$--$60\%$ of peak performance. 
For multisplit, which contributes the largest share of overall SpGEMM time among the three kernels (as shown in \autoref{fig:phases_suitesparse_H200}), we achieve approximately $40$--$50\%$ of peak performance.
These results are consistent with~\cite{gpu-multisplit}, an in-depth study of the multisplit kernel on random input streams, which reported $30$--$60\%$ of peak performance. In that study, the lower end of this range was observed for larger numbers of chunks, up to a maximum of 256.
In contrast, $g$MAGNUS supports an unbounded number of chunks while still sustaining near-optimal performance. 
For instance, wb-edu requires 2048 total chunks across two levels, yet still achieves $40\%$ and $50\%$ of peak performance for histogram and multisplit, respectively.

\begin{figure}[tbp]
\centering
\def\scale{0.75}
\includegraphics[width=\scale\linewidth]{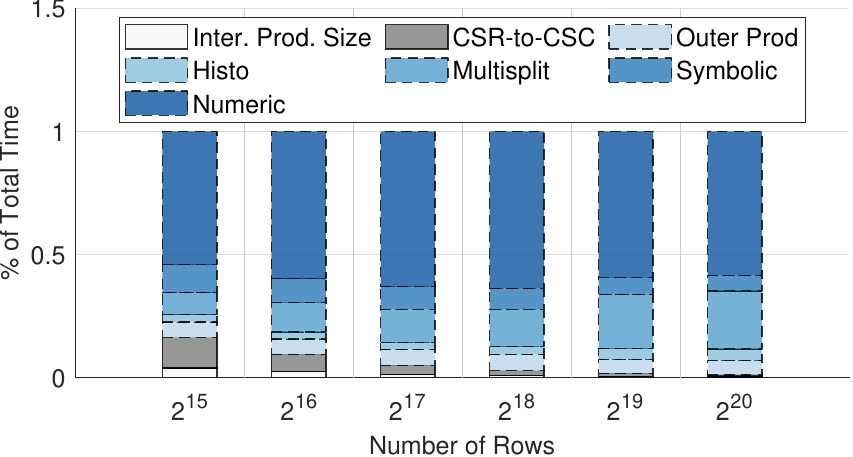}
\vspace*{-1pt}
\caption{Performance breakdown of $g$MAGNUS phases for an RMat with $A.nnz/A.n = 4$ on H200.}
\label{fig:phases_rmat}
\end{figure}

\begin{figure}[tbp]
\centering
\def\scale{0.75}
\includegraphics[width=\scale\linewidth]{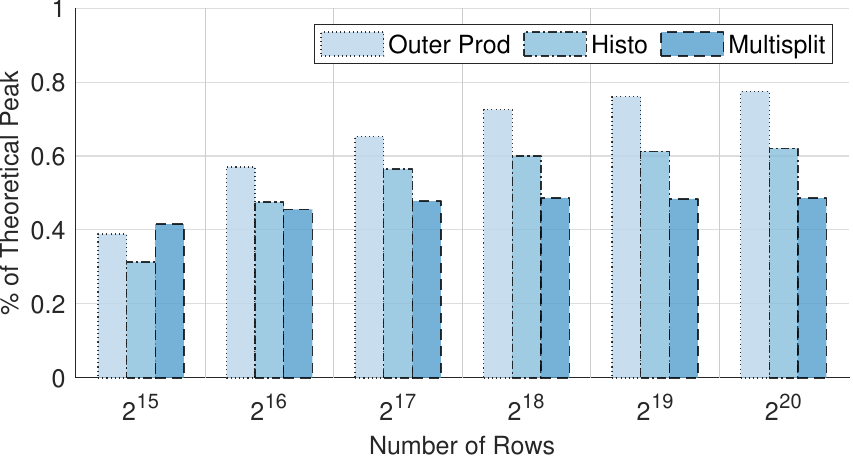}
\vspace*{-1pt}
\caption{Comparison of the core kernels in $g$MAGNUS to their theoretical upper bound for an RMat with $A.nnz/A.n = 4$ on H200.}
\label{fig:sol_rmat}
\end{figure}

\autoref{fig:sol_rmat} and \autoref{fig:phases_rmat} show the same metrics for RMats with $A.nnz/A.n=4$ and $B.nnz/B.n=8$, providing insight into performance-critical kernels as we increase the matrix dimensions. 
As the matrix scale increases, the percentage of peak performance converges to approximately 75\%, 60\%, and 50\% for the outer product, histogram, and multisplit, respectively. 
The phase breakdown shows that more time is spent in the matrix core kernels as we increase the matrix scale due to the higher numbers of heavy rows.
Additionally, the fraction of time spent in the setup phase diminishes as matrix scale increases, demonstrating that our setup overhead becomes negligible as the heavy row intermediate product sizes increase.
This convergence demonstrates that our core kernels scale excellently for our most irregular datasets, resulting in the excellent scaling of $g$MAGNUS compared to other SpGEMM algorithms.
\section{Conclusion}
This paper presents $g$MAGNUS, a novel GPU algorithm for SpGEMM designed for massive irregular matrices.
Such matrices often contain many \emph{heavy rows}, whose intermediate products exceed local memory capacity and force conventional local-memory accumulators to fall back to global-memory solutions.
The key idea behind $g$MAGNUS is to compute an intra-row reordering of the intermediate products using an optimized outer product and a hierarchical multisplit, enabling accumulation entirely in local memory.
We evaluate SYCL and CUDA implementations of $g$MAGNUS on Intel Ponte Vecchio and NVIDIA H200, respectively, using matrices from the SuiteSparse collection and recursive power-law graphs.
The results show that $g$MAGNUS outperforms five widely used baselines, including cuSPARSE and MKL, with the largest gains on massive matrices with many heavy rows.
In addition, its core kernels (outer product and multisplit) achieve near-optimal performance relative to the theoretical ``speed of light'' upper bound.


\bibliographystyle{IEEEtran}
\bibliography{refs}

\end{document}